\documentclass[useAMS,usenatbib]{mn2e}
\usepackage{amsmath}
\usepackage{amssymb}

\usepackage{graphics}
\usepackage{graphicx}
\usepackage{epstopdf}
\usepackage{footnote}
\usepackage{tablefootnote}
\usepackage{txfonts}
\usepackage{lipsum}
\usepackage{float}
\usepackage[draft]{hyperref}
\usepackage{macros}
\usepackage{fixltx2e}

\makesavenoteenv{tabular}

\DeclareSymbolFont{cmletters}{OML}{cmm}{m}{it}
\DeclareMathSymbol{v}{\mathalpha}{cmletters}{"76}

\usepackage[usenames,dvipsnames,svgnames,table]{xcolor}
\usepackage{hyperref}
\definecolor{darkblue}{rgb}{0.0,0.0,0.3}
\hypersetup{colorlinks,breaklinks,
            linkcolor=darkblue,urlcolor=darkblue,
            anchorcolor=darkblue,citecolor=darkblue}

\title[Feeding Sagittarius A*]{The Surprisingly Small Impact of Magnetic Fields On The Inner Accretion Flow of Sagittarius A* Fueled By Stellar Winds
 }
\author[S. M. Ressler, E. Quataert, J. M. Stone ]{S. M. Ressler$^{1,2},$  E. Quataert$^{1},$ J. M. Stone$^{3}$\\
$^{1}$Departments of Astronomy \& Physics, Theoretical Astrophysics Center, University of California, Berkeley, CA 94720 \\
$^{2}$Kavli Institute for Theoretical Physics, University of California Santa Barbara, Santa Barbara, CA 93107 \\
$^{3}$Department of Astrophysical Sciences, Princeton University, Princeton, NJ 08544}

\begin{document}

\maketitle

\begin{abstract}
We study the flow structure in 3D magnetohydrodynamic (MHD) simulations of accretion onto Sagittarius A* via the magnetized winds of the orbiting Wolf-Rayet stars. These simulations cover over 3 orders of magnitude in radius to reach $\approx$ 300 gravitational radii, with only one poorly constrained parameter (the magnetic field in the stellar winds).    Even for winds with relatively weak magnetic fields (e.g., plasma $\beta$ $\sim$ $10^6$),  flux freezing/compression in the inflowing gas amplifies the field to $\beta$ $\sim$ few well before it reaches the event horizon.    Overall, the dynamics, accretion rate, and spherically averaged flow profiles (e.g., density, velocity) in our MHD simulations are remarkably similar to analogous hydrodynamic simulations.  We attribute this to the broad distribution of angular momentum provided by the stellar winds, which sources accretion even absent much angular momentum transport. We find that the magneto-rotational instability is not important because of i) strong magnetic fields that are amplified by flux freezing/compression, and ii) the rapid inflow/outflow times of the gas and inefficient radiative cooling preclude circularization.  The primary effect of magnetic fields is that they drive a polar outflow that is absent in hydrodynamics.
The dynamical state of the accretion flow found in our simulations is unlike the rotationally supported tori used as initial conditions in horizon scale simulations, which could have implications for models being used to interpret Event Horizon Telescope and GRAVITY observations of Sgr A*.   
 \end{abstract}

\section{Introduction}

The accretion system immediately surrounding Sagittarius A* (Sgr A*), the supermassive black hole in the centre of The Milky Way, offers an unparalleled view of the diverse physical processes at play in galactic nuclei.  Compared to other active galactic nuclei, the luminosity of the black hole is strikingly small, only $\sim 10^{-9}$ times the Eddington limit, this places it firmly into the regime of the well-studied Radiatively Inefficient Accretion Flow (RIAF) models (see \citealt{Yuan2014} for an extensive review).  The proximity of the Galactic Centre allows for the environment immediately surrounding the black hole to be spatially resolved, including $\gtrsim$ 100s of stars in the central nuclear cluster \citep{Paumard2006,Lu2009}, the hot, X-ray emitting gas at the Bondi radius \citep{Baganoff2003}, and the ionized mini-spirals streaming inwards surrounded by the cold, molecular circumnuclear disc.  Direct constraints on the near horizon environment are now possible with the detection of several localized infrared flares orbiting the black hole within $\sim$ 10 gravitational radii ($r_g \equiv GM/c^2$, where $M$ is the mass of the black hole, $G$ is the gravitational constant, and $c$ is the speed of light) by GRAVITY \citep{GRAVITYFlare} and the first resolved mm images by the Event Horizon Telescope \citep{Doeleman2009,EHT1,EHT2} soon to come. With such a wealth of observational data, Sgr A* can be used as a test-bed of accretion models in a way that no other system can.  

It is generally believed that the black hole's gas supply is primarily set by the stellar winds of the $\sim$ 30 Wolf-Rayet (WR) stars orbiting at distances of $\sim$ 0.1-1 pc from Sgr A* \citep{Paumard2006,Martins2007,YZ2015}.  The winds shock with each other to $\sim$ keV temperatures, producing X-rays around the Bondi radius that are well resolved by \emph{Chandra}  \citep{Baganoff2003}.  However, a spherical Bondi estimate vastly over-predicts the observed Faraday rotation of the linearly polarized radio emission \citep{Agol2000,Quataert2000,Bower2003,Marrone2007}.  Instead, only a small fraction $\lesssim 10^{-3}$ of this gas reaches the horizon.  It is this material that produces the X-ray and infrared flares as well as the 230 GHz emission targeted by EHT.   

What exactly prevents most of the material at the Bondi radius from accreting is still an open debate.  Several viable models have been proposed, including those that appeal to strong outflows (\citealt{BB1999}) and those that appeal to convective instabilities that trap gas in circulating eddies \citep{Stone1999,Narayan2000,CDAF,CDBF,Pen2003}.  The range of models corresponds to a dependence of density on radius between the two extremes of $r^{-{3/2}}$ and $r^{-1/2}$, with the combination of multiple observational estimates at $\sim$ 7 different radii supporting $r^{-1}$ in the inner regions of the flow \citep{Gillessen2019} with a potential break near the Bondi radius \citep{Wang2013}.  Another key consideration is the angular momentum of the gas being fed at large radii. In the absence of magnetic fields or other processes, gas in axisymmetric flows can only accrete if it has a specific angular momentum (roughly) less than the Keplerian value at the event horizon.  
On the other hand the magnetorotational instability (MRI; \citealt{BalbusHawley}) can amplify an initially weak field 
causing gas to accrete while also driving strong magnetically dominated outflows in the polar regions.  

Most simulations of accretion onto low luminosity AGN operate either explicitly or implicitly on the assumption that 
%this conclusion from \citet{PB03B} 
the MRI is the primary driver of accretion.  For instance, General Relativistic Magnetohydrodynamic (GRMHD) simulations used to model the horizon-scale accretion flow in the Galactic Centre (e.g., \citealt{DeVilliers2003,Gammie2003,Mckinney2004,Mosci2009,
Narayan2012,Sadowski2013,Mosci2014,CK2015,EHT5}; see also \citealt{Porth2019} for a recent GRMHD code comparison) almost uniformly start from equilibrium tori (e.g. \citealt{Fishbone1976,Penna2013}) seeded with weak magnetic fields that are unstable to the MRI.  No low angular momentum gas is initially present.
In this picture, understanding the physics of the MRI and how it depends on physical parameters like the net vertical flux in the disc or numerical parameters like resolution is essential for understanding accretion physics. % in these simulations. 

Sgr A* is unique among AGN in that we can plausibly expect to directly model the accretion of gas from large radii where it is originally sourced by the winds of the Wolf-Rayet stars.  Since the hydrodynamic properties of these winds \citep{Martins2007,YZ2015} as well as the orbits of the star themselves \citep{Paumard2006,Lu2009} can be reasonably estimated from observations, the freedom in our modeling is limited mainly to the magnetic properties of the winds, which are less well known.  In principle, a simulation covering a large enough dynamical range in radius could self consistently track the gas from the stellar winds as it falls into the black hole, determining the dominant physical processes responsible for accretion and directly connecting the accretion rate, density profile, and outflow properties of the system to the observations at parsec scales.  

With this motivation, \citet{Cuadra2005,Cuadra2006,Cuadra2008} studied wind-fed accretion in the Galactic Centre with a realistic treatment of stellar winds and \citet{Cuadra2015,Russell2017} added a ``subgrid'' model to study how feedback from the black hole affects the X-ray emission. In \citet{Ressler2018} (RQS18) we built on this key earlier work by treating the winds of the WR stars as source terms of mass, momentum, and energy in hydrodynamic simulations encompassing the radial range spanning from $\sim$ 1 pc to $\sim$ $5 \times 10^{-5}$ pc ($\sim$ 300 $r_g$).  One key result of RQS18 
was that even in hydrodynamic simulations the accretion rate onto the black hole is significant and comparable to previous observational estimates (e.g., \citealt{Marrone2007,Shcherbakov2010,Ressler2017}) due to the presence of low angular momentum gas. This is in part a consequence of a coincidence that the WR stars in the Galactic Centre have winds speeds comparable to their orbital speeds, so that there is a wide range of angular momentum in the frame of Sgr A*. Another key result was that the higher angular momentum gas that could not accrete did not build up into a steady torus but was continuously being recycled through the inner $\sim$ 0.1 pc via inflows and outflows.  Because of this complicated flow structure, it is not clear what effect magnetic fields would have.  Would the rotating gas be unstable to the MRI?  Would the MRI growth time be short enough compared to the inflow/outflow time in order to significantly effect the flow structure? If so, how is the net accretion rate altered? How significant are large scale magnetic torques in transporting angular momentum? These and more are the questions we address in this work.

\citet{Ressler2019} (RQS19) presented a methodology for modeling the accretion of magnetized stellar winds by introducing additional source terms to account for the azimuthal field in each wind. In that work, we showed that a single simulation of fueling Sgr A* with magnetized winds 
can satisfy a number of observational constraints, providing a convincing argument that our model is a reasonable representation of the accretion flow in the Galactic Centre. 
First, our simulations reproduce the total X-ray luminosity observed by \emph{Chandra} \citep{Baganoff2003}, meaning that we capture at least a majority of the hot, diffuse gas at large radii. 
Second, our simulations reproduce the $r^{-1}$ density scaling inferred from observations that were taken over a large radial range \citep{Gillessen2019}, implying that we are capturing a majority of the gas at \emph{all} radii and that our inflow/outflow rates have the right radial dependence. 
Third, our simulations can reproduce the magnitude of the RM of both the magnetar (produced at $r\gtrsim 0.1$ pc, \citealt{Eatough2013}) and Sgr A* (produced at $r\lesssim $ $10^{-4}$ pc, \citealt{Marrone2007}), demonstrating that our calculated magnetic field strengths are reasonable at both small and large scales. 
Fourth, our simulations can plausibly explain the time variability of the RM of Sgr A* \citep{Bower2018},
the time variability of the magnetar's RM, as well as the time variable part of its dispersion measure \citep{Desvignes2018}. In this work, we study the dynamics of this model in more detail, with the primary focus of determining the degree to which magnetic fields alter the flow structure seen in purely hydrodynamic simulations (e.g., \citealt{Cuadra2008}, RQS18).

One key open question regarding the horizon scale accretion flow onto Sgr A* is whether or not it is magnetically arrested. This state can occur when coherent magnetic flux is consistently accreted onto the black hole and amplified by (e.g.) flux freezing \citep{Shvartsman1971} to the point that the magnetic pressure becomes large enough to halt the inflow of matter.  This configuration is generally referred to as a ``Magnetically Arrested Disc'' (MAD; \citealt{Narayan2003}).  Simulations show that accretion in the MAD state are much more time variable than their Standard and Normal Evolution (SANE) counterparts, have much stronger jets, and the bulk of accretion occurs along thin transient streams that are able to penetrate to the horizon \citep{Igumenshchev2003,Sasha2011}.  The periodicity of the polarization vectors of the localized infrared flares detected by GRAVITY favors the presence of a strong, coherent, vertical magnetic field at horizon scales \citep{GRAVITYFlare}, large enough to potentially be in a MAD state.  MAD models are also favored over SANE models of the emission from the supermassive black hole in M87 because they more naturally account for the energetics of the jet \citep{EHT5}. Simulations of magnetized wind accretion in the Galactic Centre are uniquely equipped to address the question of whether or not the winds of the WR stars can provide enough coherent magnetic flux for Sgr A* to become MAD.   

This paper is organized as follows.  \S \ref{sec:Model} reviews and summarizes the governing equations of the system including the magnetized wind source terms, \S \ref{sec:wind_test} demonstrates in an isolated stellar wind test that our method produces the desired results, \S \ref{sec:results} presents the results of 3D MHD simulations of accretion onto Sgr A*, \S \ref{sec:proga} compares and contrasts our results with previous work, \S \ref{sec:imp} discusses the implications of our work for horizon scale modeling of the Galactic Centre, and \S \ref{sec:conc} concludes.

\section{Computational Methods}
\label{sec:Model}

Our simulations use the conservative, grid-based code {\tt Athena++}\footnote{{\tt Athena++} is rewrite of the widely used {\tt Athena} code \citep{Stone2008} in the c++ language.  For the latest version of {\tt Athena++}, see https://princetonuniversity.github.io/athena/. } coupled with the model for magnetized winds outlined in RQS19.  This model is an extension of the purely hydrodynamic wind model presented in RQS18 and treats the winds of the WR stars as sources of mass, momentum, energy, and magnetic field that move on fixed Keplerian orbits.  The hydrodynamic properties of the winds are parameterised by their mass loss rates, $\dot M_w$ and their wind speeds, $v_w$.  The magnetic fields of the winds are purely toroidal as defined with respect to the spin axes of the stars and have magnitudes set by the parameter $\beta_w$, defined by the ratio between the ram pressure of the wind and its magnetic pressure at the equator (a ratio that is independent of radius in an ideal stellar wind).

\begin{figure*}
\includegraphics[width=0.45\textwidth]{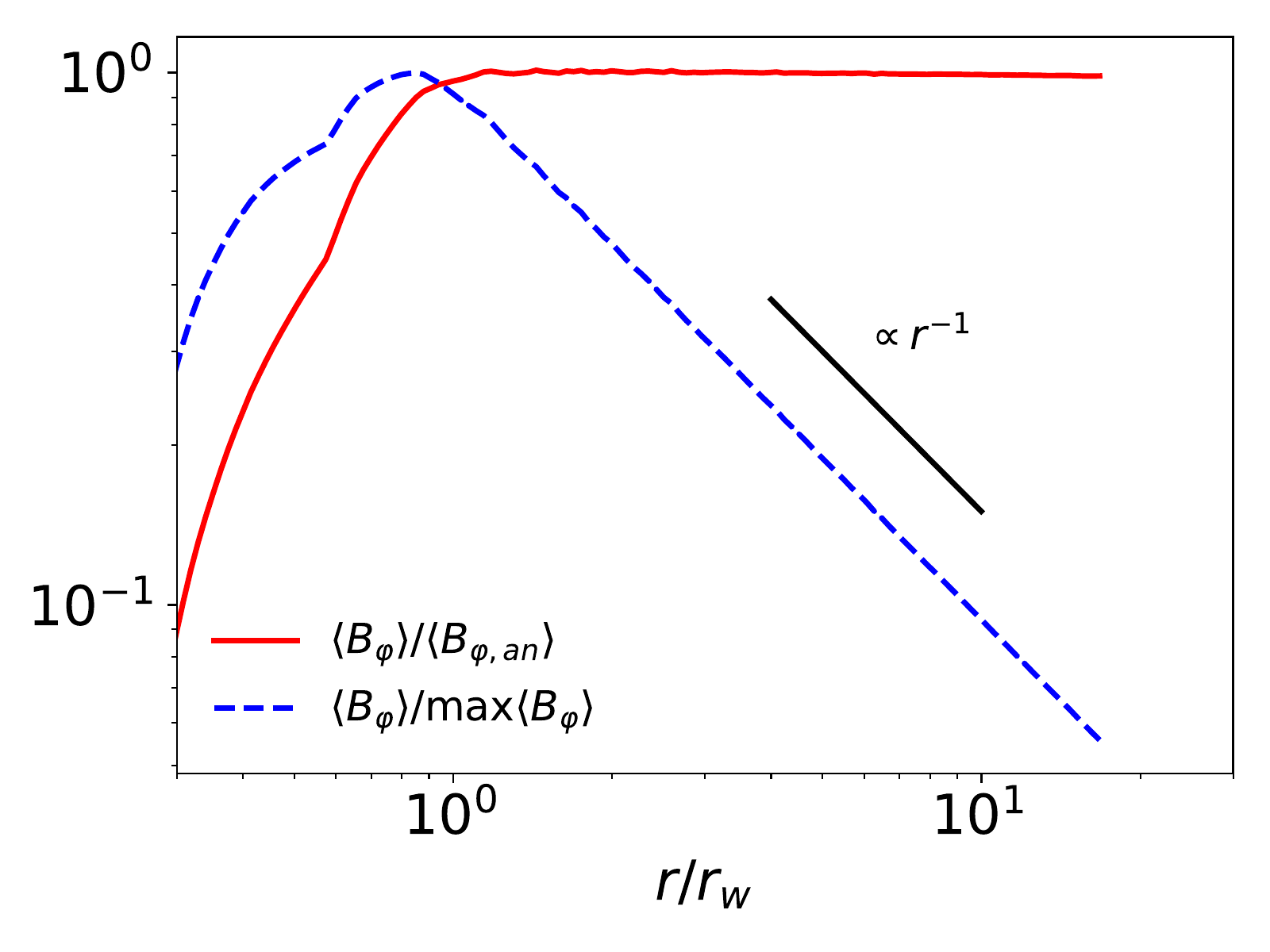}
\includegraphics[width=0.45\textwidth]{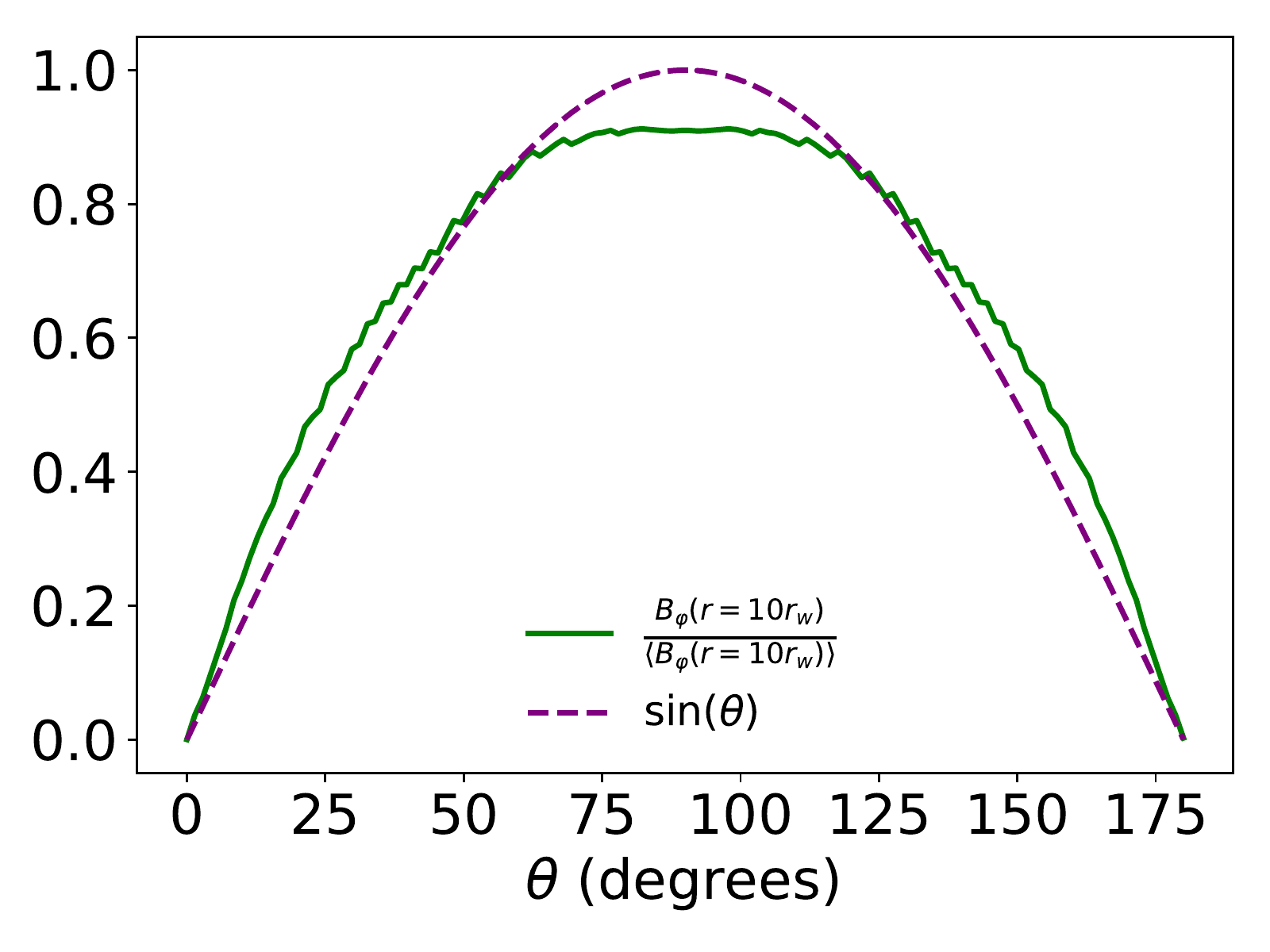}
\caption{Left: Angle averaged azimuthal magnetic field, $\langle B_\varphi \rangle$, normalized to the $r\gtrsim r_w$ analytic expectation for $\beta_w=10^2$ (solid red) and to its peak value (dashed blue).  The agreement with the analytic solution is excellent.  Right: $\theta$ dependence of $B_\varphi$ at 10 wind radii (solid green) compared to $\sin(\theta)$ (dashed purple).  The magnetic field is slightly more spread out in $\theta$ than $\sin(\theta)$ because the imbalanced magnetic pressure tends to push the gas towards the poles.  This effect is more extreme for $\beta_w =10$ (Figure \ref{fig:wind_coll}).} 
\label{fig:wind_test}
\end{figure*}

\begin{figure*}
\includegraphics[width=0.45\textwidth]{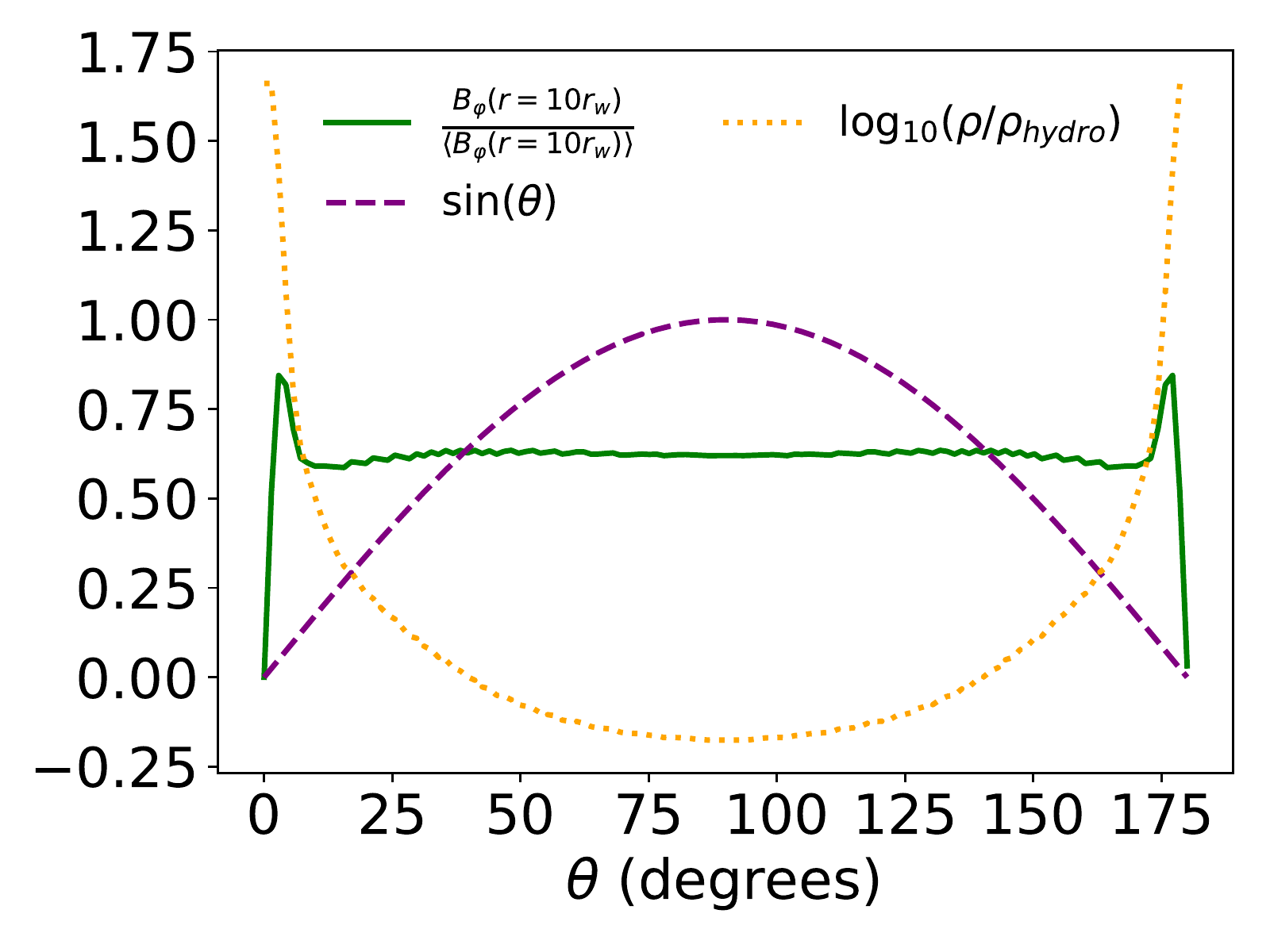}
\includegraphics[width=0.45\textwidth]{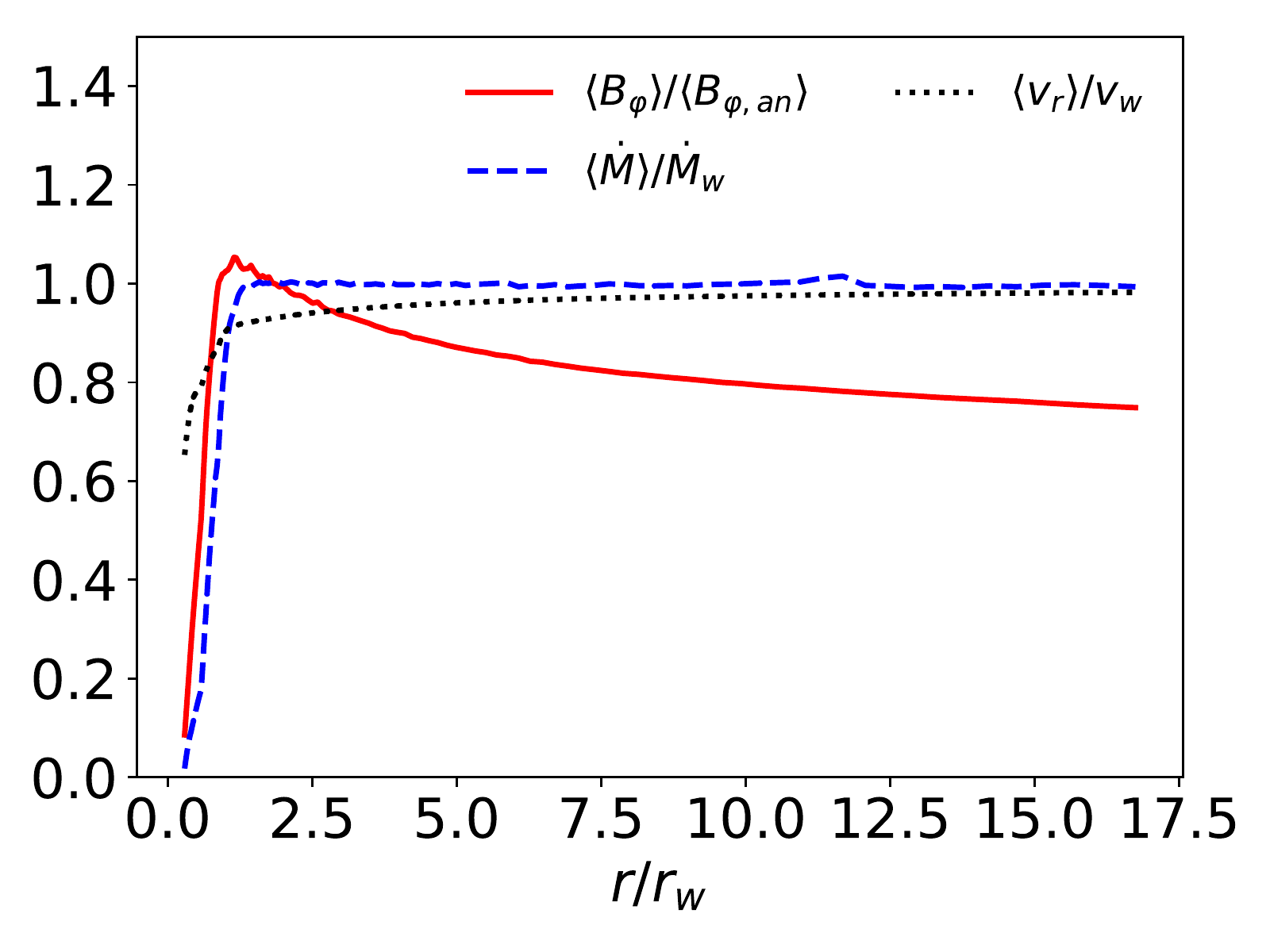}
\caption{Left: $\theta$ dependence of the azimuthal magnetic field, $B_\varphi$ (solid green), and the logarithm of the mass density divided by the purely hydrodynamic solution, $\log_{10}(\rho/\rho_{hydro})$ (dotted orange), both evaluated at 10 wind radii for $\beta_w=10$ and compared with $\sin(\theta)$ (dashed purple). Right: angle averaged $\langle B_\varphi \rangle$ (solid red), accretion rate normalized to the expected value, $\langle\dot M \rangle/\dot M_w$ (dashed blue), and radial velocity normalized to the expected value, $\langle v_r\rangle/v_w$ (dotted black). Compared to the $\beta_w=100$ case in Figure \ref{fig:wind_test}, the magnetic field is now strong enough to collimate the wind, enhancing the density by almost a factor of 100 at the poles.  Despite this, the net accretion rate and wind speed are still consistent with the input parameters. Nonetheless, we focus on $\beta_w \ge 100$ for our simulations to avoid the collimating effect of magnetic fields on the stellar winds.}
\label{fig:wind_coll}
\end{figure*}

In brief, the equations solved in our simulations are
\begin{equation}
\label{eq:cons}
\begin{aligned}
\frac{\partial \rho}{\partial t} + \mathbf{\nabla} \cdot \left(\rho \mathbf{v}\right) &= f \dot \rho_{w} \\
\frac{\partial \left(\rho \mathbf{v}\right)}{\partial t} + \mathbf{\nabla} \cdot \left(P_{tot} \mathbf{I} + \rho \mathbf{v}\mathbf{v} - \frac{\mathbf{B}\mathbf{B}}{4\pi}\right) &= -\frac{\rho GM_{BH}}{r^2} \hat r   \\&+ f \dot \rho_{w} \langle {\mathbf v_{w,net}}\rangle \\
\frac{\partial \left(E \right)}{\partial t}  + \mathbf{\nabla} \cdot \left[(E+P_{tot})\mathbf{v} - \mathbf{v}\cdot \mathbf{B} \mathbf{B}\right] &= - \frac{\rho G M_{BH}}{r} \mathbf{v}\cdot \hat r + \langle \dot E_B \rangle \\ &+ \frac{1}{2} f \dot \rho_{w} \langle { |\mathbf v_{w,net}}|^2\rangle - Q_{-} \\
\frac{\partial \mathbf{B}}{\partial t} - \mathbf{\nabla} \times \left( \mathbf{v}\times \mathbf{B}\right) &=\mathbf{\nabla} \times  \left(  \tilde E_w\right),
\end{aligned}
\end{equation} 
where $\rho$ is the mass density, $\mathbf{v}$ is the velocity vector, $\mathbf{B}$ is the magnetic field vector, $E = 1/2 \rho v^2 + P/(\gamma-1) + B^2/(8\pi)$ is the total energy, $\gamma=5/3$ is the non-relativistic adiabatic index of the gas, $P_{tot} = P +  B^2/(8\pi)$ is the total pressure including both thermal and magnetic contributions, $Q_{-}$ is the cooling rate per unit volume due to radiative losses caused by optically thin bremsstrahlung and line cooling (using $Z= 3 Z_\odot$ and $X=0$), $f$ is the fraction of the cell by volume contained in the wind, $\dot \rho_{w}  = \dot M_{w} / V_{w}$, $V_{w} = 4\pi /3$ $ r_{w}^3$, ${\mathbf v_{w,net}}$ is the wind speed in the fixed frame of the grid, $\langle\rangle$ denotes a volume average over the cell, $\dot E_B$ is the magnetic energy source term provided by the winds, and $\tilde E_w$ is the average of the wind source electric field, $\mathbf{E_w}$, over the appropriate cell edge (see Equations 22-24 of \citealt{Stone2008}). Each `wind' has a radius of $r_w \approx$ 2 $\sqrt{3} $ $\Delta x$, where $\Delta x$ is the edge length of the cell containing the centre of the star.

\section{Isolated, Magnetized Stellar Wind Test}
\label{sec:wind_test}
To test that our implementation of the source terms drives a magnetized wind with the desired properties, we place a stationary wind in the centre of a 3D, 1 pc$^3$ grid and run for 4 wind crossing times.  The mass-loss rate of the wind is $\dot M_w = 10^{-5}$ $M_\odot$/yr, the wind speed is $v_w = 1000 $ km/s, and radiative cooling is disabled.  We choose $\beta_w = 100$ to ensure that the magnetic field is non-negligible but relatively weak.
\begin{figure*}
\includegraphics[width=0.95\textwidth]{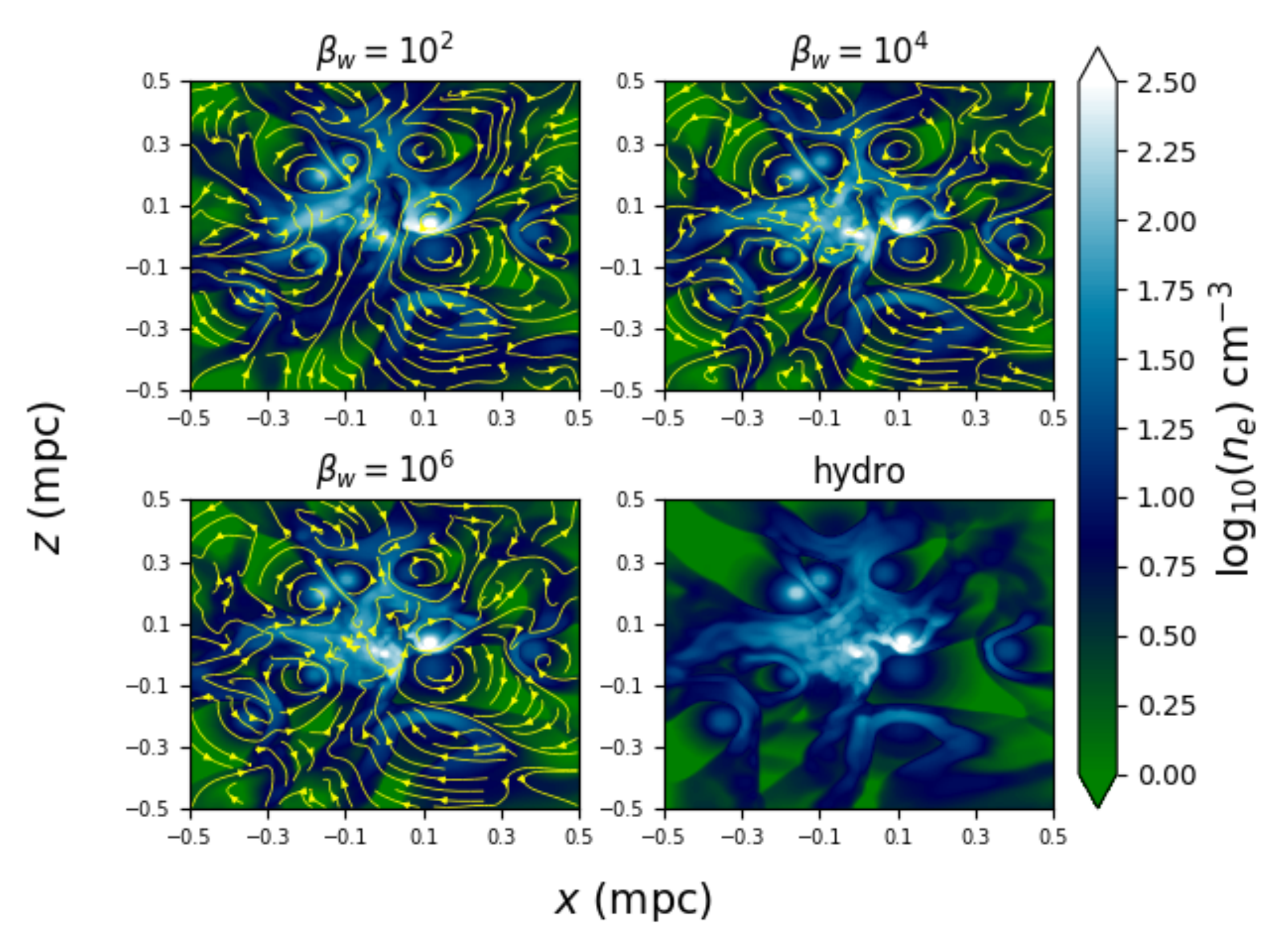}
\caption{2D slice in the plane of the sky of electron number density overplotted with projected magnetic field lines for the inner $\sim$ 0.5 pc of our $\beta_w = 100$ (top left), $\beta_w=10^4$ (top right), $\beta_w=10^6$ (bottom left), and hydrodynamic (bottom right) simulations. Each `star' in our simulation provides a purely toroidal magnetic field with direction determined by the random, independently chosen spin axes of the stars. No significant difference is seen in the simulations at this scale because the magnetic fields are relatively weak compared to the ram and thermal pressures of the gas.
 }
\label{fig:rho_outer}
\end{figure*}

\begin{figure*}
\includegraphics[width=0.85\textwidth]{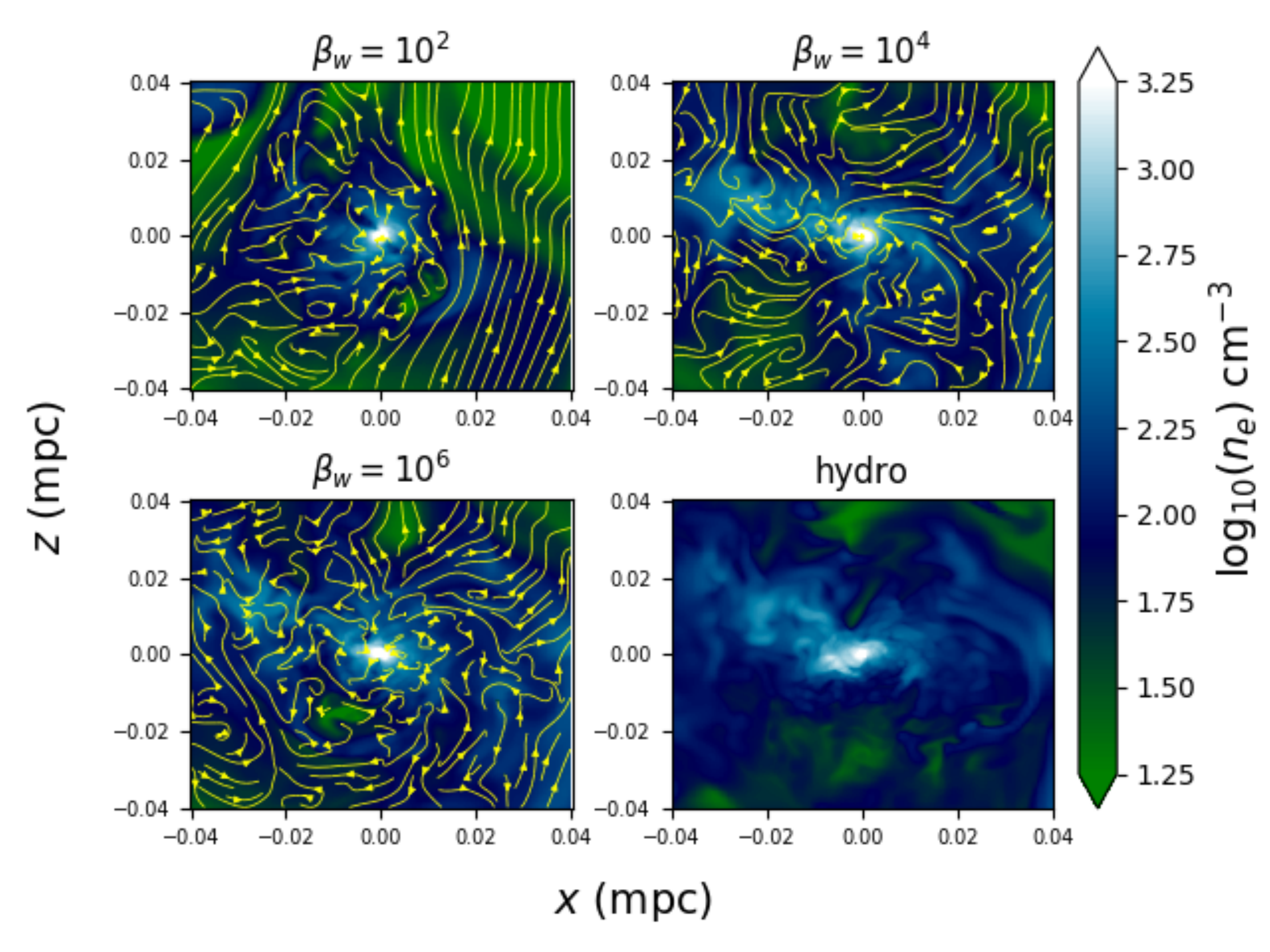}
\includegraphics[width=0.85\textwidth]{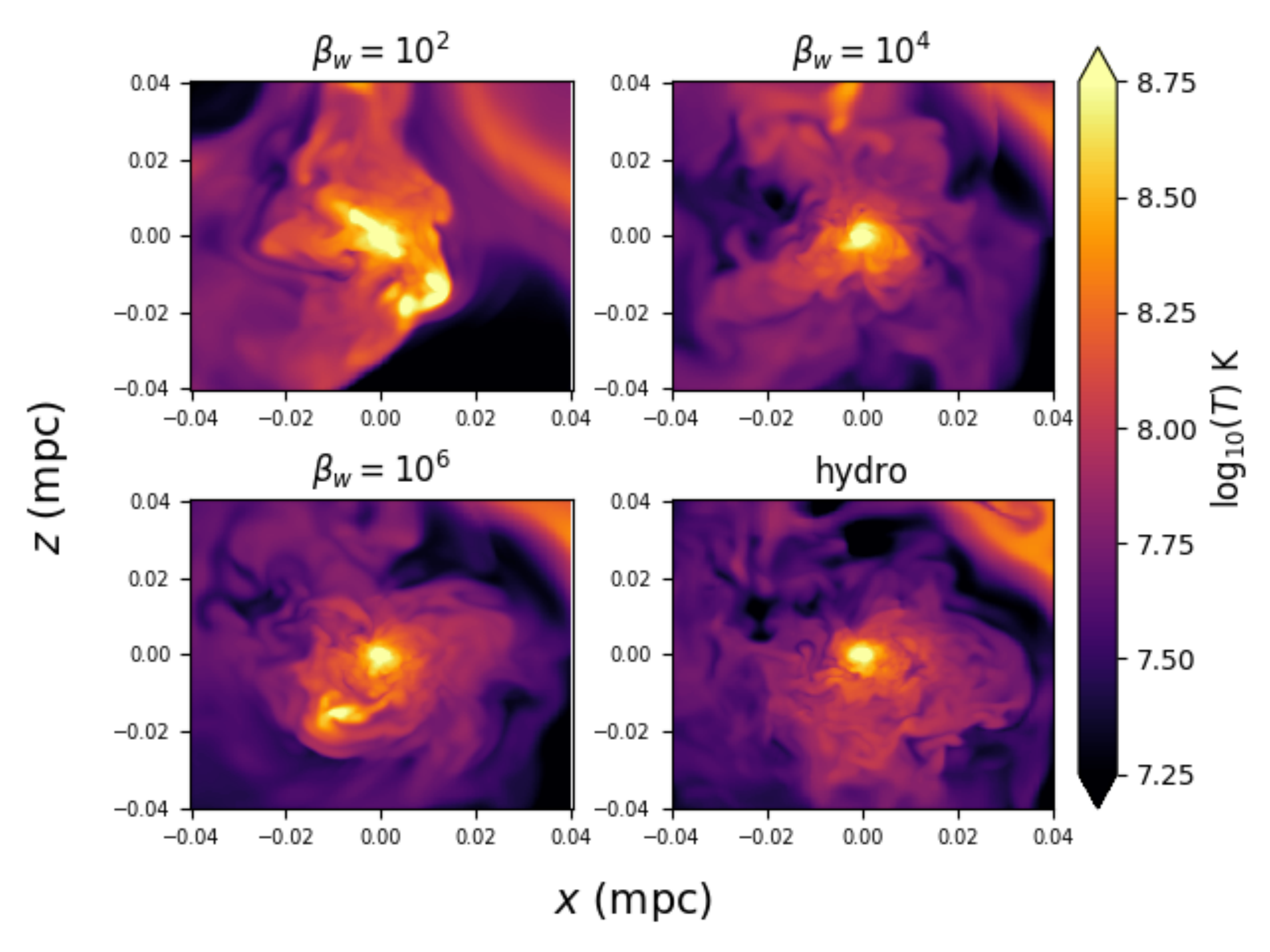}
\caption{2D slice in the plane of the sky of electron number density overplotted with projected magnetic field lines (top four panels) and temperature (bottom four panels) for the inner $\sim$ 0.05 pc of our simulation. Compared to the 0.5 pc scale in Figure \ref{fig:rho_outer}, the magnetic fields in the $\beta_w=10^2$ and $\beta_w=10^4$ simulations are more dynamically important and thus clear differences are seen in the density distribution compared to the hydrodynamic case. In addition, the larger the field strength in the winds, the larger the spatial scale over which the field lines are coherent.  
}
\label{fig:rho_inner}
\end{figure*}

The left panel of Figure \ref{fig:wind_test} shows that the angle-averaged $\varphi$ component of the magnetic field matches the analytic expectation, scaling as $r^{-1}$ as determined by flux conservation. The other components of the field are negligible.  The right panel of Figure \ref{fig:wind_test} shows the dependence of $B_\varphi$ on polar angle $\theta$ at a distance of 10 times the wind radius.  Since the $\varphi$ source term in the induction equation is $\propto$ $\sin(\theta)$, a dynamically unimportant magnetic field would also be $\propto \sin(\theta)$.  For $\beta_w =100$, however, corresponding to a magnetic pressure that is $1$\% of the ram pressure, the unbalanced $P_m \propto \sin(\theta)^2$ pushes the gas away from the midplane and towards the poles.  This leads to the field being slightly lower than prescribed in the midplane and slightly higher near the poles, by a factor of $\lesssim 10 \%$.  We emphasize that this result is not an error in our model but a self-consistent consequence of magnetic stresses in the wind, which tend to collimate the flow \citep{Sakurai1985}.  However, it is also important to note that the $\sin(\theta)$ dependence of the source term in the induction equation was chosen simply because it vanishes at 0 and $\pi$ and not based on detailed modeling of the angular structure of MHD winds.  

The angular structure of the wind seen in Figure \ref{fig:wind_coll} becomes even more pronounced for $\beta_w=10$, where the magnetic pressure is now $10 \%$ of the ram pressure.  Here the wind becomes highly collimated, as shown by the left panel of Figure \ref{fig:wind_coll}, where the density is now concentrated at the poles and the magnetic field is roughly independent of polar angle.  At the same time, the total mass outflow rate and angle averaged wind speeds are still in good agreement with the intended $v_w$ and $\dot M_w$ as shown in the right panel of Figure \ref{fig:wind_coll}.   

As noted in RQS19, when $\beta_w $ is further decreased to $\lesssim 5$ the magnetic pressure becomes large enough to accelerate the wind and make the solution inconsistent with the input parameters.  Because of this we limit our studies to $\beta_w \gtrsim 10$ and focus primarily on $\beta_w \ge 10^2$.  We show later that our results are insensitive to $\beta_w$ for $\beta_w \sim$ $10^2-10^6$.

\section{3D Simulation of Accreting Magnetized Stellar Winds Onto Sgr A*}
\label{sec:results}

\subsection{Computational Grid and Boundary/Initial Conditions}

We use a base grid in Cartesian coordinates of $128^3$ covering a box size of (2 pc)$^3$ with an additional 9 levels of nested static mesh refinement.\footnote{Note that in RQS18 and RQS19 it was stated that the box size of our simulations was (1 pc)$^3$.  This was an error.  The box size in both works was actually (2 pc)$^3$, as it is here.}  This doubles the effective resolution every factor of $\sim$ 2 decrease in radius so that the length of an edge of the smallest cubic cell is $\approx$ $3 \times 10^{-5}$ pc. No additional mesh refinement is used (e.g.) near the stellar wind source terms.  The inner boundary of our simulation is approximately spherical, with a radius equal to twice the length of an edge of the smallest cubic cell, corresponding to $r_{in} \approx 6 \times 10^{-5}$ pc$ \approx 1.6 \times 10^{-3 \prime\prime} $ $\approx 300$ $r_g$.  All cells with centre points within this radius are set to have zero velocity and floored density/pressure, yet the magnetic field is allowed to freely evolve.  Even though we generally expect the solution just outside this radius in our simulations to have large inwards radial speeds, we chose to set the velocity to zero inside $r_{in}$ for simplicity and have found that it does not strongly affect our results.   First, we have tested that simulations using this inner boundary condition are able to successfully reproduce a spherical Bondi flow even when the sonic radius is smaller than $r_{in}$ (that is, when gas within the inner boundary radius is in causal contact with gas at larger radii), and second, the simulations presented in RQS18 using the same inner boundary showed that $-v_r$ just outside the inner boundary was still able to reach the local sound speed (see Figure 11 in that work).   The outer boundary of each of our simulations is set at the faces of the computational box using ``outflow'' conditions, where primitive variables are simply copied from the nearest grid cell into the ghost zones. 

Our simulations use the Harten-Lax-van Leer+Einfeldt (HLLE; \citealt{Einfeldt1988}) Riemann solver and 2nd order piece-wise linear reconstruction on the primitive variables.

For the WR stars, we use the orbits, mass loss rates, and wind speeds exactly as described in RQS18, drawing primarily from \citet{Martins2007}, \citet{Cuadra2008}, \citet{Paumard2006}, and \citet{Gillessen2017}. These values differ slightly from those used in RQS19, where we modified the mass loss rates and wind speeds of four stars (within reasonable systematic observational uncertainties) to show that our simulations could reproduce the observed RM of the Galactic Centre magnetar.      Each wind is given a randomly chosen direction for its spin axis that determines the azimuthal direction for the magnetic field; this random selection is made only once for each star so that each simulation we run has the same set of spin axes. Note that RQS19 used a different set of spin axes, but we have found our results insensitive to this choice.  Here we also use a value of $r_{in}$ that is a factor of 2 smaller than RQS19. We ran a total of 5 simulations; four in MHD with $\beta_w = 10,10^2,10^4$ and $10^6$, and one in hydrodynamics (i.e., $\beta_w \rightarrow \infty$).

In \S \ref{sec:wind_test} we showed that stellar winds with $\beta_w=10$ in our model become highly collimated.  We have found that this collimation has nontrivial effects on the resulting dynamics of the inner accretion flow (in particular, altering the angular momentum direction at small radii) in a way that makes separating the effects of large magnetic fields from this extra hydrodynamic consideration difficult.  Furthermore, the precise nature of this collimation depends on our choice of angular dependence of $\tilde E_w$ in Equation \eqref{eq:cons}, which was arbitrary.  Thus we do not find it instructive to include $\beta_w=10$ in our analysis, though we note that the main conclusions derived from our $\beta_w=10^2-10^6$ simulations are consistent with those derived from the $\beta_w=10$ simulations that we have run.

We initialize each simulation with floored density and pressure, zero velocity, and no magnetic field, starting at 1.1 kyr in the past.  Here, for consistency with RQS18  January 1, 2017 is defined as the present day, i.e., $t=0$.  The simulations are run for 1.3 kyr to a final time of $t_f = 200$ yr.  In Appendix \ref{app:t0_comp} we argue that our results are independent of the arbitrary choice of starting our simulations 1.1 kyr in the past by comparing with simulations that start 9 kyr in the past.

Finally, we use floors on the density and pressure (see RQS18), and a ceiling on the Alfven speed (which is effectively an additional floor on density; see RQS19).

To ensure that our results are well converged, we ran an additional simulation that used a factor of 4 finer resolution within $\sim$ 0.06 pc, though with a shorter total run time.  As shown in Appendix \ref{app:resolution}, we find that our simulations show no significant dependence on resolution.

\subsection{Overview}
Figure \ref{fig:rho_outer} shows a 1 pc$^2$ 2D slice in the plane of the sky (centred on the black hole) of the electron number density over-plotted with magnetic field lines for $\beta_w=10^2,10^4,$ and $10^6$ compared to the hydrodynamic case.  Magnetic fields do not significantly alter the dynamics at this scale because even for $\beta_w=100$ the magnetic pressure in the winds is insignificant compared to their ram pressure.  Thus, all panels are nearly identical.  Slices of the temperatures show similarly small differences from the right panel of Figure 7 in RQS18 and are thus not included here.  Figure \ref{fig:rho_outer} shows that, as desired, the magnetic fields lines wrap around the ``stars,'' which show up as dense circles typically surrounded by bow shocks.  Again, since the field is not dynamically important at this scale the field lines for different $\beta_w$'s all have essentially the same geometry.

The top four panels of Figure \ref{fig:rho_inner} again show 2D slices of the electron number density overplotted with field lines but on a scale of $\sim$ 0.08 pc, ten times smaller than Figure \ref{fig:rho_outer}.  The bottom four panels of the same figure show 2D slices of temperatures. While the $\beta_w=10^6$ run still looks similar to the hydrodynamic simulation, the $\beta_w=10^4$ and (particularly) $\beta_w=100$ runs show significant differences.  This is because, as we shall show, the field in the latter cases starts to become dynamically important at this scale.   The field lines become increasingly ordered with decreasing $\beta_w$, going from mostly tangled for $\beta_w=10^6$ (where the field is easily dragged along with the flow) to mostly coherent for $\beta_w=100$ (where the field can resist the gas motion).  This will have important implications for the field geometry at small radii in \S \ref{sec:field_structure}.

\begin{figure*}
\includegraphics[width=0.95\textwidth]{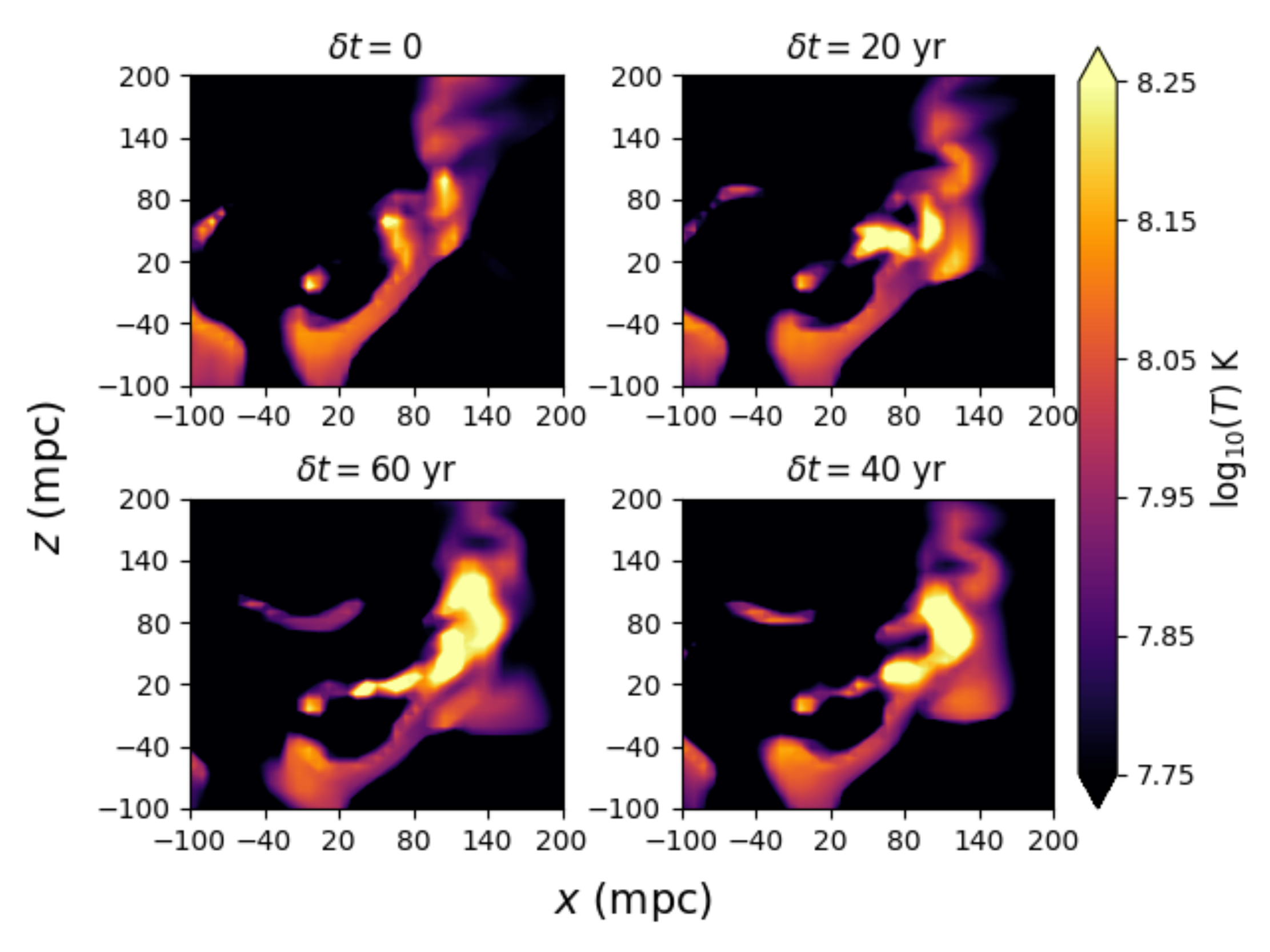}
\caption{Time series of jet formation over the course of 150 years in the $\beta_w=10^2$ simulation. Plotted are 2D gas temperature slices in the plane of the sky for the inner $\sim$ 0.04 pc; note that we only show the upper right quadrant of the simulation to highlight the jet. Here $\delta t$ is the time elapsed since the first snapshot. Time proceeds clockwise starting from the upper left panel.  The color scale differs from that used in Figure \ref{fig:rho_inner} and was chosen to particularly highlight the highest temperatures. As time progresses, a collimated, high temperature outflow emerges asymmetrically from small radii until it reaches $r \sim$ 0.3 pc.  This `jet' is present only sporadically during the course of the $\beta_w=10^2$ simulation and not at all in the higher $\beta_w$ simulations.
}
\label{fig:jet_time_series}
\end{figure*}

In the $\beta_w=10^2$ simulation alone, a prominent large scale, hot, collimated outflow can be seen at particular times (at $t=0$ in Figure \ref{fig:rho_inner} it is relatively weak, though can be seen reaching to $\sim$ 0.01 pc below the black hole in the top left temperature panel).  Figure \ref{fig:jet_time_series} presents a time series of the gas temperature, highlighting $T\ge 6 \times 10^7$ K and spanning $t = -540$ yr to $t = -480$ yr in 20 yr increments. In the initial frame, no clear outflow structure is seen, only strong shocks between winds. As time progresses, however, a thin $\ge 10^8$ K outflow appears coming out from the right side of the black hole, $\sim$ parallel to the angular momentum direction at this time.  This outflow is magnetically driven and originates at small radii, as we shall show in the next section.

\begin{figure}
\includegraphics[width=0.45\textwidth]{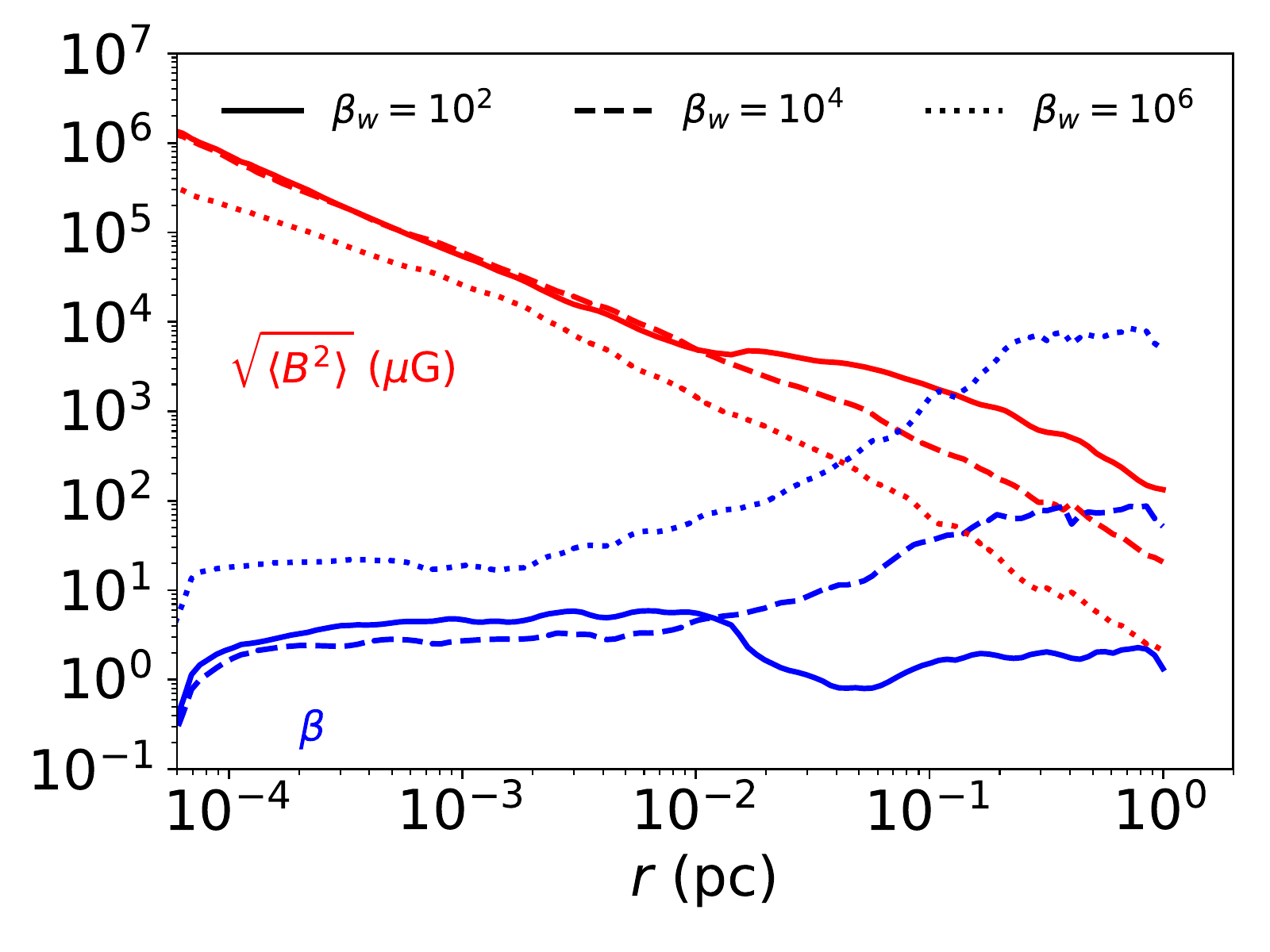}
\caption{Comparison between the root-mean-squared magnetic field strength (red), $\sqrt{\langle B^2 \rangle}$, and plasma $\beta$ (blue), $\langle P\rangle / \langle P_m \rangle$, for different values of $\beta_w$, which quantifies the magnetization of the WR stellar winds. Even though the field strength varies by 2 orders of magnitude at large radii (corresponding to a 4 orders of magnitude difference in $\beta$), the field strengths at small radii are all within a factor of $\lesssim$ 2 ($\beta$'s within a factor of $\lesssim$ 3).  This is because the field tends to be compressed and amplified by being dragged along with the gas motion until $\beta$ reaches $ \sim$ a few.  }
\label{fig:beta_comp}
\end{figure}

Figure \ref{fig:beta_comp} shows the angle-averaged root-mean-squared (rms) magnetic field strength and plasma $\beta \equiv P/P_m$, where $P_{m}$ is the magnetic pressure, as a function of radius for different values of $\beta_w$.  As expected, at large radii ($\gtrsim 0.1$ pc), the rms field and $\beta$ scale simply as $\sqrt{1/\beta_w}$ and $\beta_w$, respectively.  At small radii ($\lesssim 10^{-2}$ pc), however, there is a much weaker dependence of the rms field and $\beta$ on $\beta_w$.  In fact, both $\beta_w=10^4$ and $\beta_w=10^2$ reach $\beta \sim 2$ and $\sim$ 1 G field strengths by $10^{-4}$ pc.  Even in the $\beta_w=10^6$ case, the field strength ($\beta$) at small radii is only a factor of $\sim$ 2 less (3-4 larger) than in the $\beta_w=10$ simulation.  This is why RQS19 found that the rotation measure of Sgr A* and the net vertical flux threading the inner boundary of the simulation were roughly independent of $\beta_w$, since both quantities are set by the field at the innermost radii. Though this result might seem like a clear signature of a field regulated by the magnetorotational instability (MRI), we argue in \S \ref{sec:MRI} that this is not the case, and that instead the amplification is due to flux freezing in the inflow.

\begin{figure}
\includegraphics[width=0.45\textwidth]{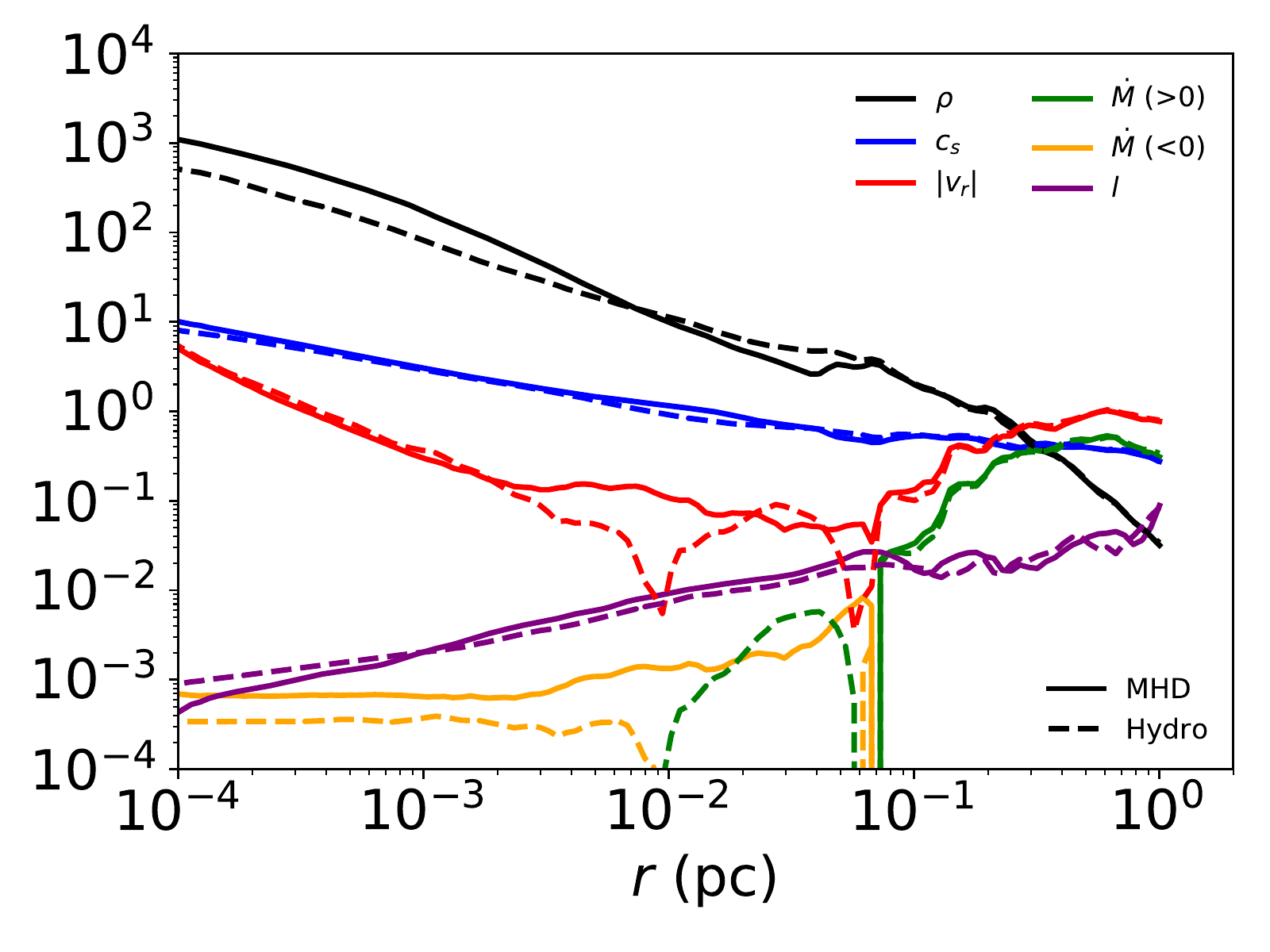}
\includegraphics[width=0.45\textwidth]{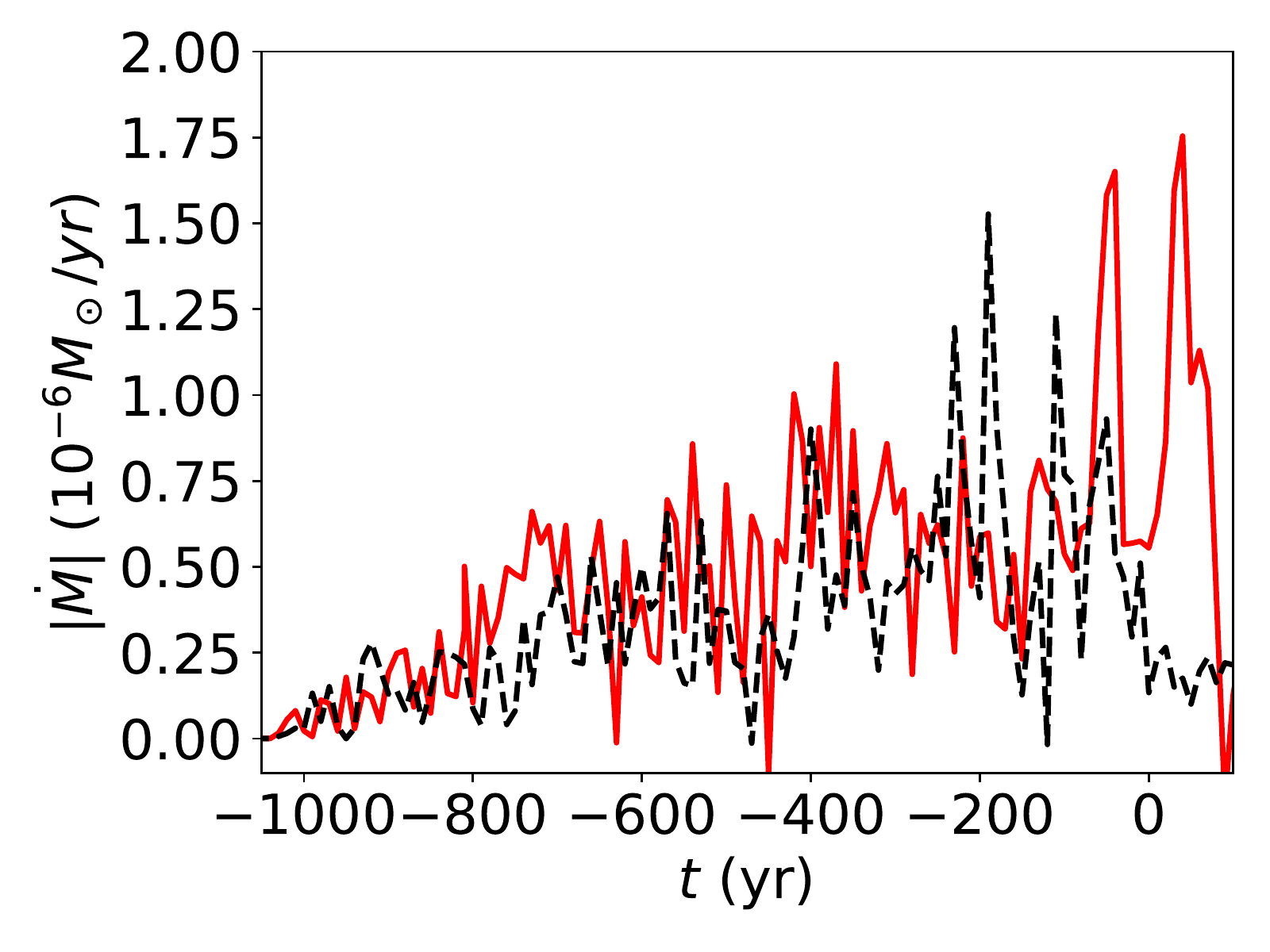}
\caption{Comparison between the $\beta_w=10^2$ MHD simulation (solid) and a purely hydrodynamic simulation (dashed).  Top: present day angle-averaged mass density, $\rho$ (M$_\odot$/pc$^3$), sound speed, $c_s$ (pc/kyr), radial velocity, $|v_r|$ (pc/kyr), specific angular momentum, $l$ (pc$^2$/kyr), and net mass accretion rate, $\dot M$ (M$_\odot$/kyr), as a function of distance from the black hole. Positive net accretion rates are green, while negative net accretion rates are orange. Bottom: Mass accretion rate as a function of time measured at $\approx$ 2 mpc $\approx$ 9700 $r_g$.  Despite the relatively large magnetization of the stellar winds, the magnetic field has an almost negligible effect on the radial profiles. The small difference in density (and hence, accretion rate) is caused by the slightly different time dependence of the accretion rate leading to a different realization of the flow at $t=0$ even though the statistics in time are similar. These conclusions are independent of $\beta_w$.  }
\label{fig:mhd_comp}\end{figure}

Despite the clear morphological differences at the $\sim$ 0.08 pc scale in the density/temperature (Figure \ref{fig:rho_inner}) and the fact that the flow can reach $\beta$ of $\sim$ a few over orders of magnitude in radius (Figure \ref{fig:beta_comp}), the radially averaged gas properties in the MHD simulations remain strikingly similar to the hydrodynamic results at all radii even for $\beta_w=100$. This is shown in Figure \ref{fig:mhd_comp}, which shows the angle and time averaged density, sound speed, radial velocity, angular momentum, and accretion rate in addition to the accretion rate through the inner boundary as a function of time for both the $\beta_w=100$ and the hydrodynamic run. Here and throughout we refer to the ``accretion rate'' as the \emph{net} accretion rate including both inflow and outflow components, i.e., $\langle \dot M \rangle = \langle 4 \pi \rho v_r r^2\rangle$. Though there can be as large as a factor of three difference in accretion rate (corresponding to a difference in density at small radii) at specific times, on average, the accretion rate through the inner boundary is unchanged by the presence of the magnetic fields, falling between $\sim$ 0.25 and 1.5 $\times 10^{-6}$ $M_{\odot}/yr$.\footnote{Due to the chaotic nature of our simulations, the instantaneous value of the accretion rate at $t=0$ is not as robust as the time-averaged value.}  The differences in the average sound speed and radial velocity are negligible.  We have tested that this result also holds for different values of the inner boundary radius. 

The net accretion rates shown in Figure \ref{fig:mhd_comp} (and in all our simulations) are negative and roughly constant in radius from the inner boundary out to $r\sim$ $10^{-2}$ pc.  Then it rises in magnitude between $r \sim$ $10^{-2}$ pc and $r \sim 10^{-1}$ pc with a sign that fluctuates with time.  Finally for $r\gtrsim 10^{-1}$ pc, it is positive and increasing with radius. A net accretion rate that is constant in radius is expected for a flow in steady-state in the absence of source terms. Our simulations, however, have a time-variable source of mass describing the contributions of stellar winds, depending on the stellar wind properties and stellar locations, the latter of which change as the stars proceed along their orbits.  In the limit of a large number of stars, this time dependence can be small if at each radial distance from the black hole there are always a similar number of stellar winds (or at least a similar total mass-loss rate).  This is roughly the case for the stellar winds in our simulations for $10^{-1}$ pc $\lesssim$ $r$ $\lesssim$ $0.4$ pc, where a majority of the stars are located.  Thus the source term in mass for $10^{-1}$ pc $\lesssim$ $r$ $\lesssim$ $0.4$ pc is roughly constant in time, and by $t=0$ a steady state is reached with positive accretion rate that increases with increasing radius.  On the other hand, at any given time, only a handful of the closest approaching stars lie between $6 \times 10^{-2}$ pc $\lesssim$ $r$ $\lesssim$ $10^{-1}$ pc so the source term in mass is time variable in this region.  Because of this, the regions between $10^{-2}$ pc $\lesssim$ $r$ $\lesssim$ $10^{-1}$ pc never reach a steady state but instead depend on the time-dependent location of the stars and the properties of their winds, both of which are observationally constrained.  For reference, the mass-weighted inflow time, $r/\langle v_r (v_r<0)\rangle_\rho$, is shorter than the simulation run time for all radii $\lesssim 0.1$ pc and shorter than a third of the simulation run time for all radii $\lesssim$ 0.06  pc, so that in the absence of source terms most of the gas between $10^{-2}$ pc $\lesssim$ $r$ $\lesssim$ $10^{-1}$ pc would have reached inflow equilibrium.  The time-dependence of the location of the stellar winds,  however, results in the magnitude of the accretion rate in this region increasing with radius though temporally fluctuating in sign.   For smaller radii, however, with $r\lesssim$ $10^{-2}$ pc, there are no significant source terms and the dynamical time is short compared to the time-scale for the temporal variability of the stellar winds sourcing the flow
so that the flow reaches a negative accretion rate that is constant with radius.  Finally, the angle-averaged flow properties at $r\gtrsim 0.4$ pc (outside most of the stellar winds) approach those of a steady Parker wind \citep{Parker1965}, but our simulations are not run long enough to fully reach this steady state.  Since our focus is on the inner accretion flow, however, this is not a concern.

\begin{figure}
\includegraphics[width=0.45\textwidth]{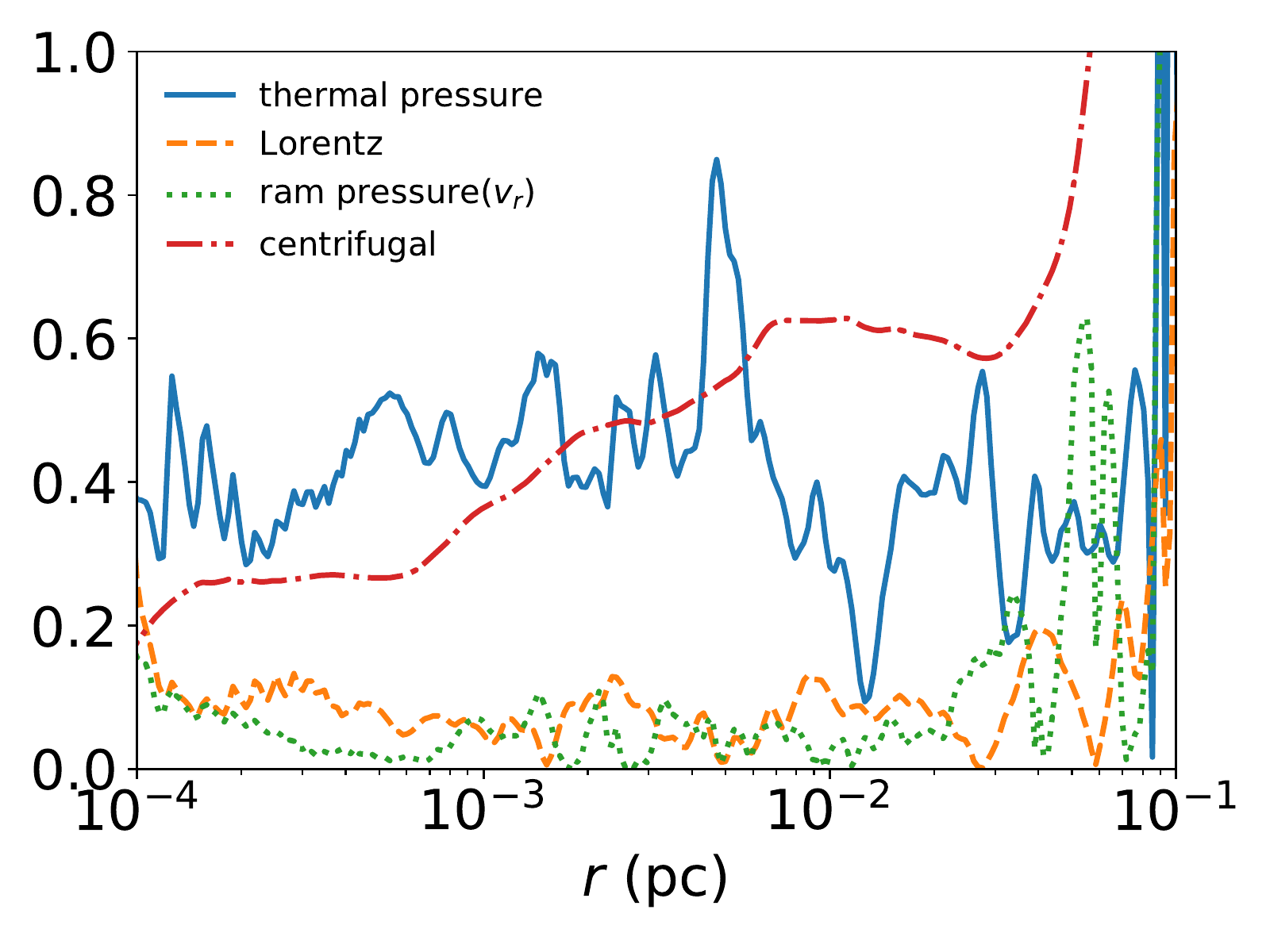}
\caption{Various components of the radial force exerted on a parcel of gas relative to the gravitational force, $\rho G M/r^2$, as a function of radius in our $\beta_w=10^2$ simulation.  Plotted are the angle and time-averaged radial component of the pressure gradient (solid), $-\partial P/\partial r$, Lorentz force (dashed), $\hat r \cdot[ (\mathbf{\nabla} \times \mathbf{B} )/(4 \pi)] \times \mathbf{B} $, the $v_r$ portion of the advection derivative (dotted), $-\rho v_r\partial v_r/\partial r$, and centrifugal force (dot-dashed), $\rho v_\varphi^2/r$.  The Lorentz force is $\approx$ 10\% of the gravitational force for most radii, comparable to the ``ram pressure force'' of the $v_r$ component of the fluid velocity.  Thermal pressure and rotation each balance about 40\% of gravity, providing a majority of the radial support.
 }
\label{fig:press_radius}
\end{figure}

To help understand why MHD and hydrodynamic simulations display only small differences in the angle-averaged radial profiles of fluid quantities (Figure \ref{fig:mdot_beta_comp}), Figure \ref{fig:press_radius} shows the various time and angle-averaged components of the outwards radial force balancing gravity for our $\beta_w=10^2$ simulation.  This includes the thermal pressure force, the Lorentz force, the centrifugal force, and the radial ram pressure force.  The magnetic field accounts for only $\lesssim$ 10\% of the total force, with the thermal pressure and centrifugal forces accounting for $\sim$ 40\% each and the radial ram pressure force accounting for $\sim$ 10\%. So although $\beta$, which takes into account only thermal pressure, is $\sim$ 2 at small radii for this simulation (Figure \ref{fig:beta_comp}), the effect of the magnetic field is reduced because of the large centrifugal and ram pressure contributions.   

\subsection{Dynamics of The Inner Accretion Flow}
\label{sec:dynamics}
To facilitate analysis of the accretion flow at small radii it is useful to define time intervals over which the angular momentum vector of the gas is relatively constant in time. Due to the stochastic nature of the simulations, this occurs at different times for each run, often not centred at $t=0$.  The purpose of this analysis, however, is to understand the general properties of the accretion disc, outflow, and magnetic field structure, not to make overly specific predictions for the present day. We expect that the intervals we choose are representative of the general accretion flow dynamics and structure. 

\begin{figure}
\includegraphics[width=0.45\textwidth]{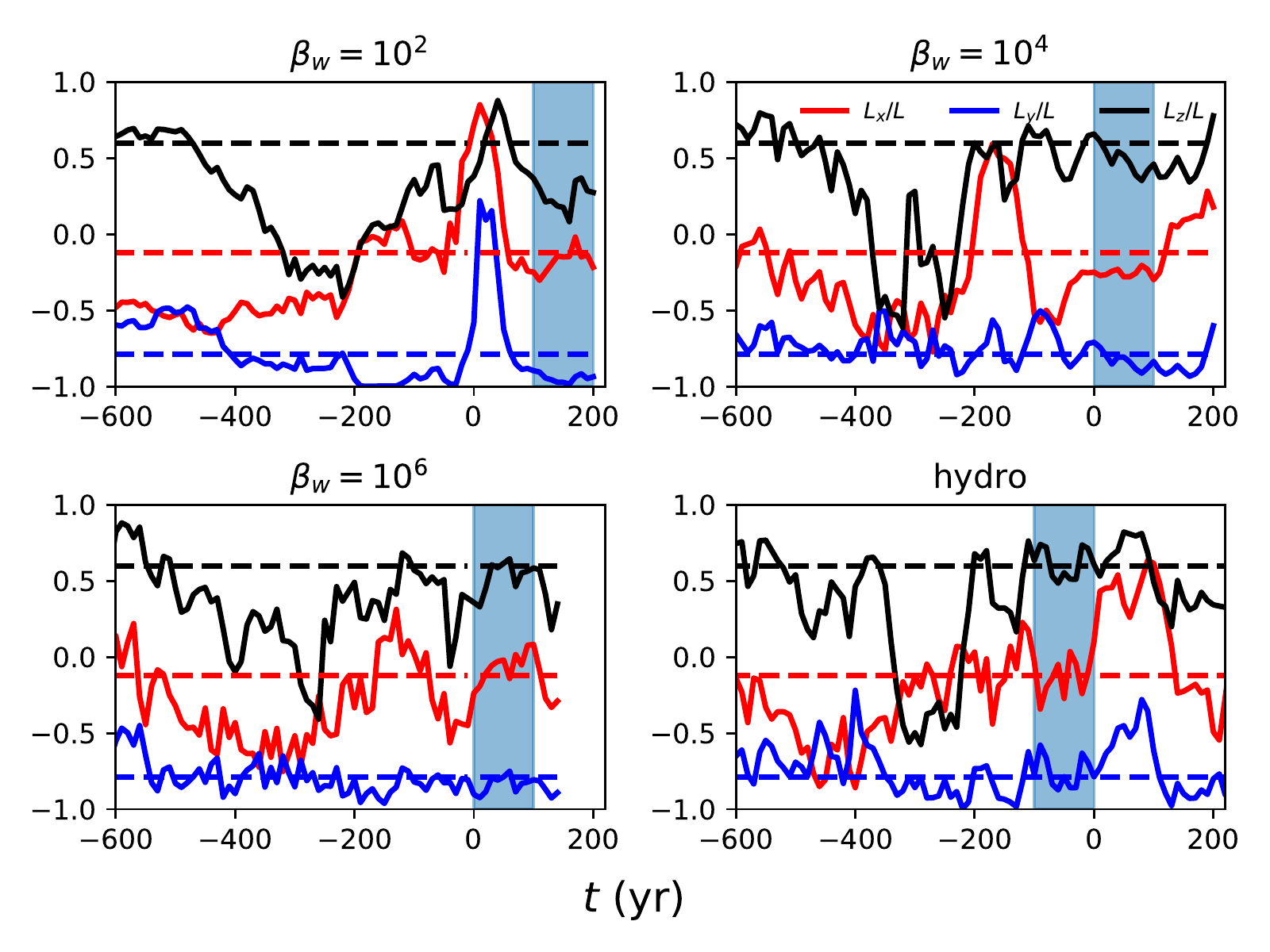}
\caption{Angular momentum direction as a function of time for the gas in our $\beta_w=10^2,10^4,10^6$ and hydrodynamic simulations, averaged in radius and angle over $5 \times 10^{-4}$ pc to $3 \times 10^{-2}$ pc.  The blue shaded regions represent the time intervals that we choose to analyze the inner accretion flow, over which the angular momentum vector is relatively stable. Dashed lines represent the three components of the angular momentum direction vector of the clockwise stellar disc \citep{Paumard2006,Lu2009}.  }
\label{fig:L_time_interval}\end{figure}

Figure \ref{fig:L_time_interval} shows the three components of the angle and radius-averaged (over the interval $r = 5 \times 10^{-4}$ pc and $r = 3 \times 10^{-2}$ pc) angular momentum direction vector as a function of time for our four simulations.  We use this information to choose our particular choice of time intervals for averaging the flow structure:  $ [100,200]$ yrs, $[0,100]$ yrs, $[0,100]$ yrs, and $[-100,0]$ yrs for $\beta_w = 10^2,10^4,10^6$ and the hydrodynamic simulation, respectively.  All of these intervals have angular momentum directions that are approximately constant in time and nearly aligned with the stellar disc containing about half of the WR stars.  The angular momentum of the accretion flow is aligned with that of the stellar disc most of the time, though for the $\beta_w=10^2$ simulation it has more frequent and larger deviations from the stellar disc than in the hydrodynamic simulation.  The most significant of these is seen near $t=0$, where the angular momentum of the gas in the $\beta_w=10^2$ simulation is nearly anti-aligned with the stellar disk for a brief $\sim$ 50 yr period. 
Note that the \emph{magnitude} of the angle and time-averaged angular momentum is similar for all simulations, being $\sim$ 0.5 $l_{kep}$ for $r\lesssim 0.1$ pc (see the left panel of Figure 14 in RQS18; the angle and time-averaged $l$ in MHD differs at most by 20\% from that in hydrodynamics as shown in Figure \ref{fig:mhd_comp}).  

\begin{figure*}
\includegraphics[width=0.95\textwidth]{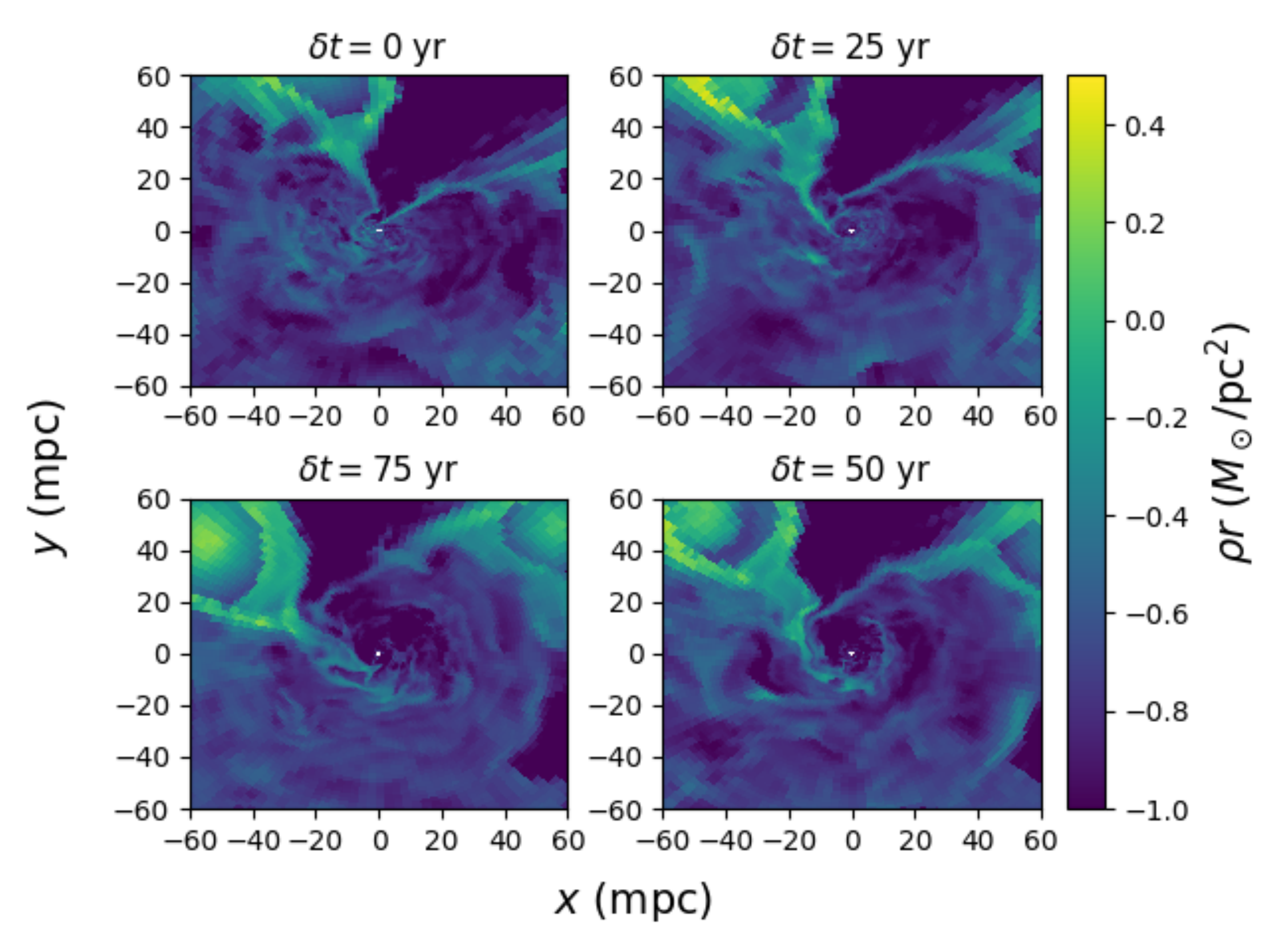}
\caption{Sequence of midplane slices of the mass density weighted by radius separated by 25 year intervals for the hydrodynamic simulation. Here $\delta t$ is the time elapsed since the first snapshot and time proceeds clockwise starting from the upper left.  Material provided by the nearest two stellar winds (E20/IRS 16C in the upper left quadrant and E23/IRS 16SW in the upper right quadrant) streams inward but mostly has too much angular momentum to accrete without any redistribution of angular momentum.  Instead, the streams of material ultimately hit a centrifugal boundary and then ``spray'' outwards on the opposite side of the black hole from which they approached.  The bulk of the gas does not circularise nor form a steady disc.  
}
\label{fig:time_series_hydro}
\end{figure*}

\begin{figure*}
\includegraphics[width=0.95\textwidth]{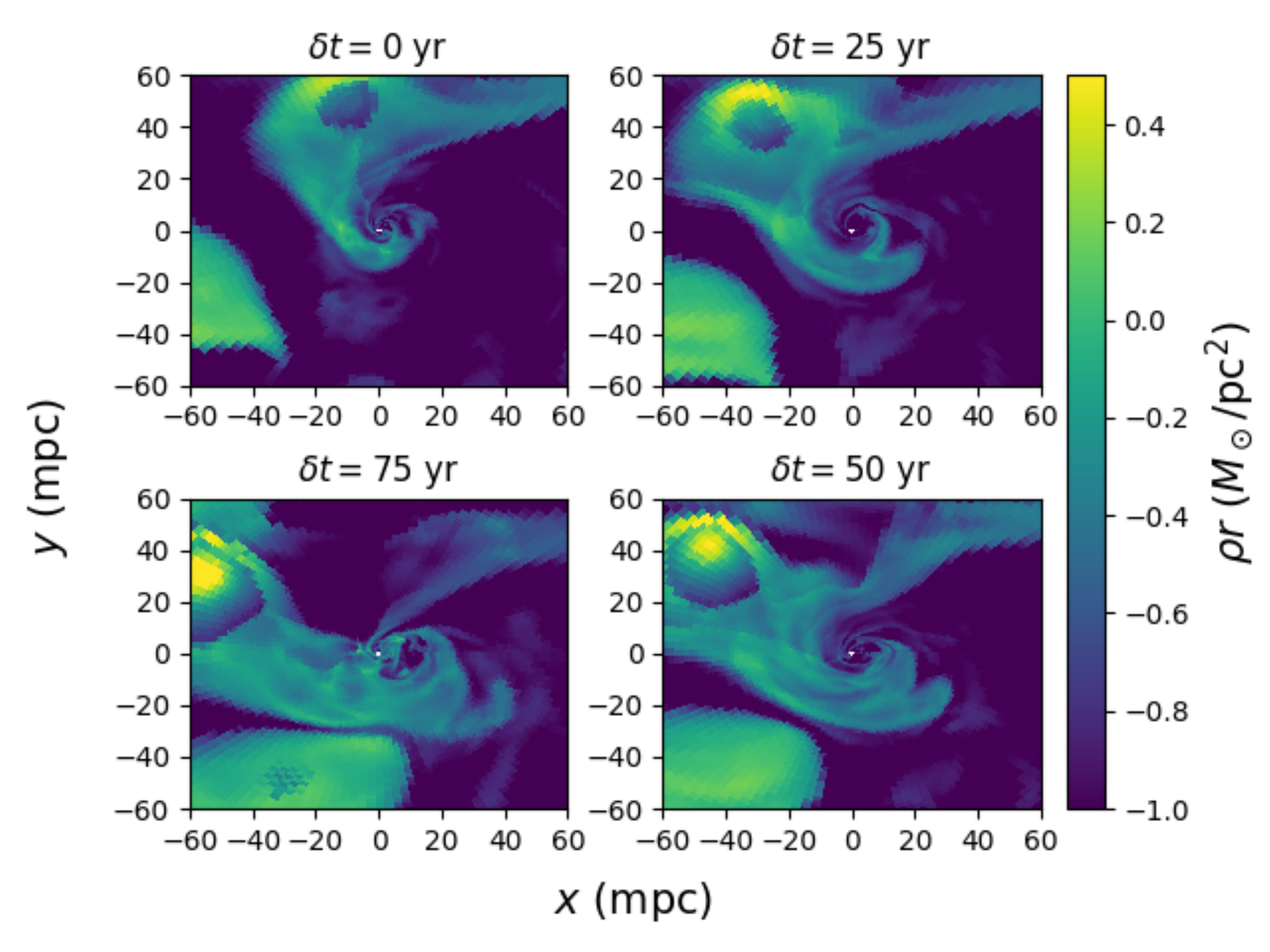}
\caption{Sequence of midplane slices of the mass density weighted by radius separated by 25 year intervals for the $\beta_w=10^2$ simulation. Time proceeds clockwise starting from the upper left panel. Note that $\delta t=0$ is at a different absolute time relative to Figure \ref{fig:time_series_hydro}.  Instead of simply streaming in and ``spraying'' outwards as seen in the hydrodynamic case (Figure \ref{fig:time_series_hydro}), strong magnetic fields are able to redirect the outflowing, high angular momentum gas towards the polar regions so that the midplane slice pictured here is mostly comprised of spiraling inflow. As in the hydrodynamic simulation, gas does not truly circularise into a disc but either accretes or outflows after only $\lesssim$ a few orbits around the black hole. While magnetic fields do provide a non-negligible torque that can remove angular momentum from the gas, this torque has limited time to operate and does not significantly modify the accretion rate, which is very similar in the hydrodynamic and MHD simulations (see Figure \ref{fig:mhd_comp}).  The two stellar winds providing most of the material in this plot are E20/IRS 16C in bottom left quadrant and E23/IRS 16SW in the upper left quadrant. 
}
\label{fig:time_series_beta_1e2}
\end{figure*}

\begin{figure*}
\includegraphics[width=0.95\textwidth]{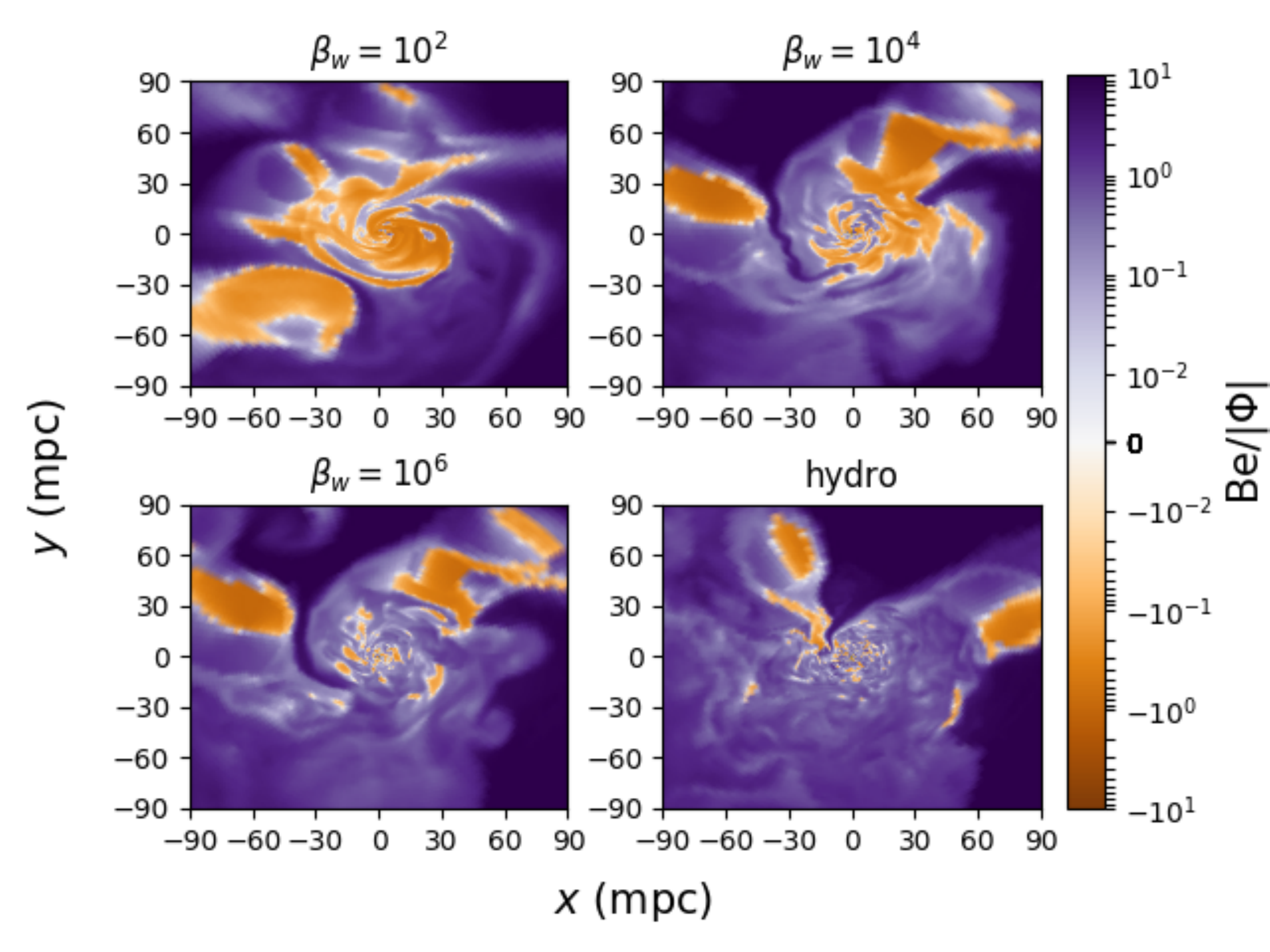}
\caption{Midplane contours of the Bernoulli parameter, Be$\equiv|\mathbf{v}|^2/2 + \gamma/(\gamma-1) P/\rho - GM/r$, divided by the gravitational potential, $|\Phi| = GM/r$,  in the four different simulations at representative times. Orange denotes bound material while purple denotes unbound material.  Absent magnetic fields, the relatively high angular momentum gas provided by the nearby stellar winds is mostly unbound with too much angular momentum to accrete (see also Figure \ref{fig:time_series_hydro}).  Strong magnetic fields (as present in the $\beta_w=10^2$ and $10^4$ simulations), however, can torque the gas enough that it loses some angular momentum and becomes moderately bound to the black hole.  The energy released by this process drives polar outflow.}
\label{fig:bernoulli_comp}
\end{figure*}

Defining a new $`z'$ direction as the direction of the time averaged angular momentum vector, Figure \ref{fig:time_series_hydro} and Figure \ref{fig:time_series_beta_1e2} show time series of the midplane ($\theta=\pi/2$) mass density on $\sim$ 0.1 pc scales, weighted by radius (see Figure \ref{fig:mhd_comp}) in the hydrodynamic and $\beta_w=10^2$ simulations, respectively.  A time series for the $\beta_w=10^4$ ($\beta_w=10^6$) simulation is not shown but it looks qualitatively very similar to the $\beta_w=10^2$ (hydrodynamic) case.  Figure \ref{fig:bernoulli_comp} shows the Bernoulli parameter, a measure of how bound the gas is to the black hole, in the same frame for all four simulations at a representative time.  These figures show that the majority of the unbound, high angular momentum gas in the midplane at large radii is provided by the closest one or two stellar winds (namely, those of E23/IRS 16SW and E20/IRS 16C) as they orbit the black hole in all simulations.  As this material streams inwards, however a clear difference is seen in the behavior at smaller radii in the different runs. In the hydrodynamic case, each fluid element largely conserves its angular momentum and energy, thus remaining unbound. Gravity is only strong enough to bend the inflowing streams of gas around the black hole until they emerge on the other side as a spray of outflow that sends the gas out to larger radii without much accretion.  The high angular momentum gas does not spend enough time at small radii to circularize or form a disc; instead the supply of matter at small radii is continually being lost and replenished.  The same is true of weakly magnetized simulations (i.e., $\beta_w=10^6$ and higher).

In MHD with strong magnetic fields (i.e., $\beta_w=10^2$ and $\beta_w=10^4$), however, this picture is different.  Now the strong fields ($\beta\sim$ a few at small radii, Figure \ref{fig:beta_comp}) are able to efficiently remove some angular momentum and energy from the gas via large-scale torques.  This results in the originally unbound material becoming bound as it falls in so that its trajectory alters to form an inward spiral that ultimately accretes instead of spraying out the other side to large radii.  The main difference, however, as we shall argue in \S \ref{sec:interp}, is that the outflow present in the midplane of the hydrodynamic simulation is now redirected to the polar regions.  As in the hydrodynamic case, the gas with high angular momentum does not spend enough time at small radii to circularize or form a true disc. This is because it generally accretes (after being subjected to magnetic torques) or is dumped into an outflow before completing even a few orbits.  Thus in both cases, the gas supply at small radii is continually being recycled and is set mostly by the hydrodynamic properties of the winds, in particular the wide range of angular momentum produced by the stellar winds.

\begin{figure*}
\includegraphics[width=0.95\textwidth]{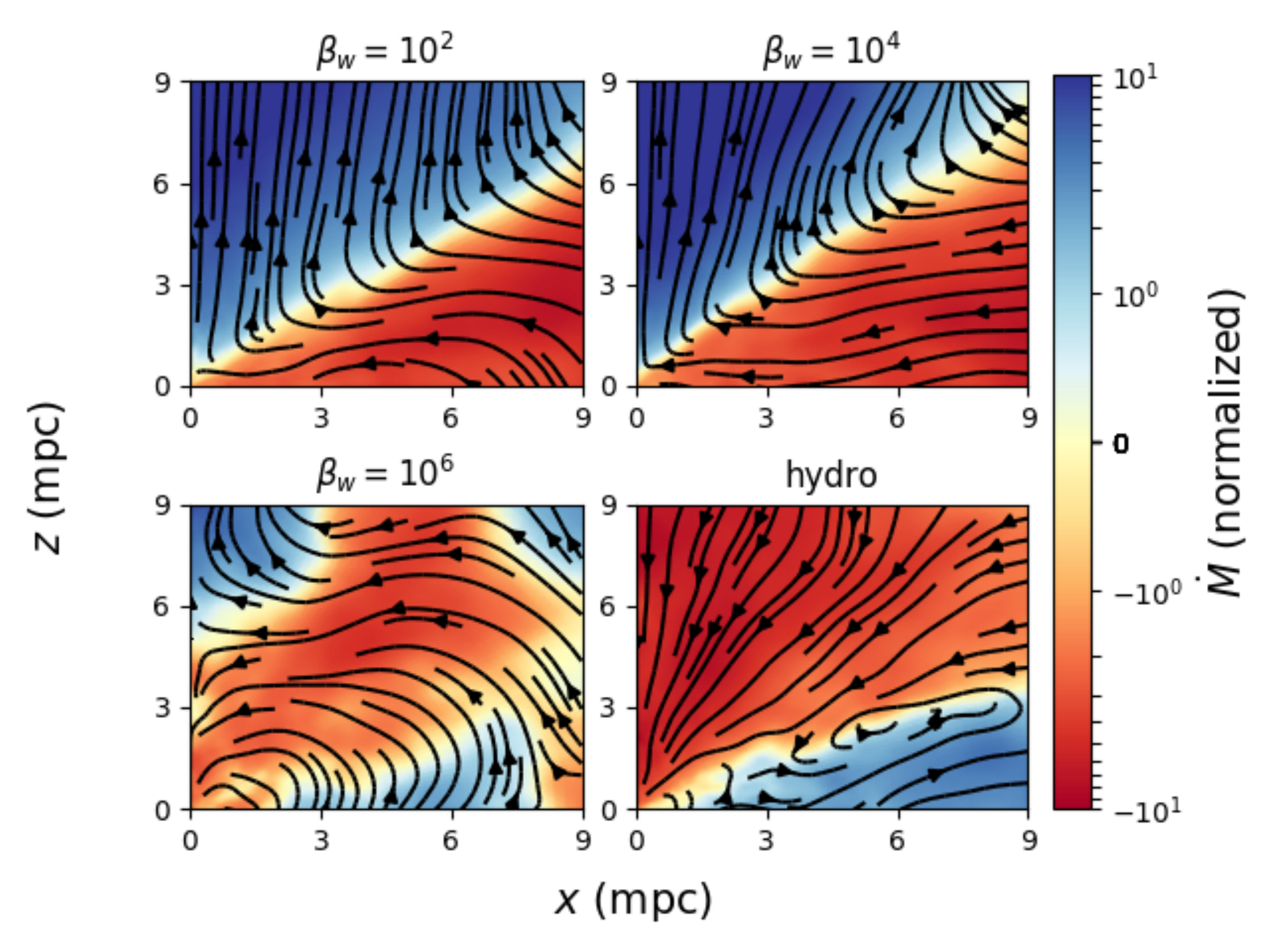}
\caption{Time and $\varphi$-averaged accretion rate  on mpc scales for each simulation, where the $z$ direction is defined as the net angular momentum of the gas (Figure \ref{fig:L_time_interval}). Red represents inflow, blue represents outflow, and the accretion rate has been folded over the midplane and normalized such that the absolute value at 5 mpc is unity. For sufficiently large magnetic fields, the inflow/outflow structure seen in the hydrodynamic case is reversed, because the field is strong enough to redirect the outflow and confine it to the polar regions. }     
\label{fig:mdot_beta_comp}\end{figure*}

\begin{figure*}
\includegraphics[width=0.95\textwidth]{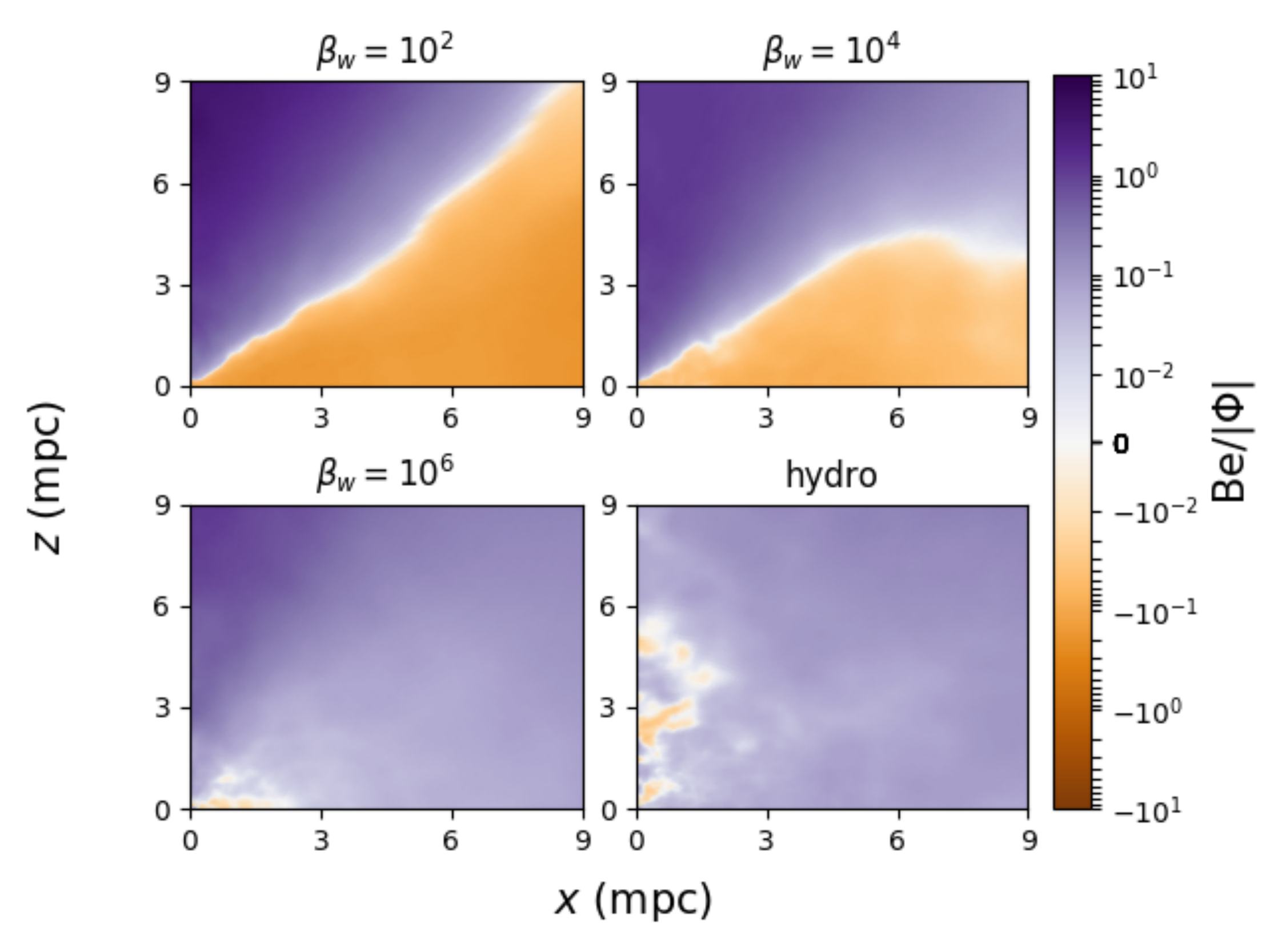}
\caption{Time and $\varphi$-averaged Bernoulli parameter, Be, normalized to the gravitational potential, $|\Phi| = GM/r$ on mpc scales for each simulation, where the $z$ direction is defined as the net angular momentum of the gas (Figure \ref{fig:L_time_interval}). Orange represents bound, purple represents unbound, and the Bernoulli parameter has been folded over the equator.  Without magnetic fields, the material is slightly unbound throughout the domain except for some slightly bound material near the polar axis.  Magnetic fields provide torque, releasing energy from the high angular momentum gas in the midplane and depositing it in the polar outflow (Figure \ref{fig:mdot_beta_comp}). }     
\label{fig:bernoulli_beta_comp}\end{figure*}

Focusing now on the poloidal structure of the flow, Figure \ref{fig:mdot_beta_comp} shows the $\varphi$ and time-averaged accretion rates for our four simulations while Figure \ref{fig:bernoulli_beta_comp} shows the same for the Bernoulli parameter.  The hydrodynamic midplane structure described above results in a net outflow of high angular momentum, modestly unbound material in the midplane, while low angular momentum material freely falls in along the poles. The polar inflow also contains some higher angular momentum, unbound (Be/$|\Phi| \lesssim 10^{-2}$ but $\ge 0$) material that eventually hits a centrifugal barrier and turns aside and adds to the midplane outflow.   For $\beta_w=10^6$, where the field is relatively weak, the same structure is seen.  However, for $\beta_w=10^4$ and $10^2$, the hydrodynamic accretion structure is completely reversed. For these more magnetized flows, not only is there net inflow of bound, Be/$|\Phi| <$ 0, material in the midplane, but the energy released from the gas as it loses angular momentum due to magnetic torques is deposited into an unbound, Be/$|\Phi| \sim $ 10 polar outflow.  As evidenced by the fact that the net accretion rates are comparable in both cases, this outflow is similar to the one present in the hydrodynamic simulation but redirected from the midplane to the poles.  

\begin{figure}
\includegraphics[width=0.45\textwidth]{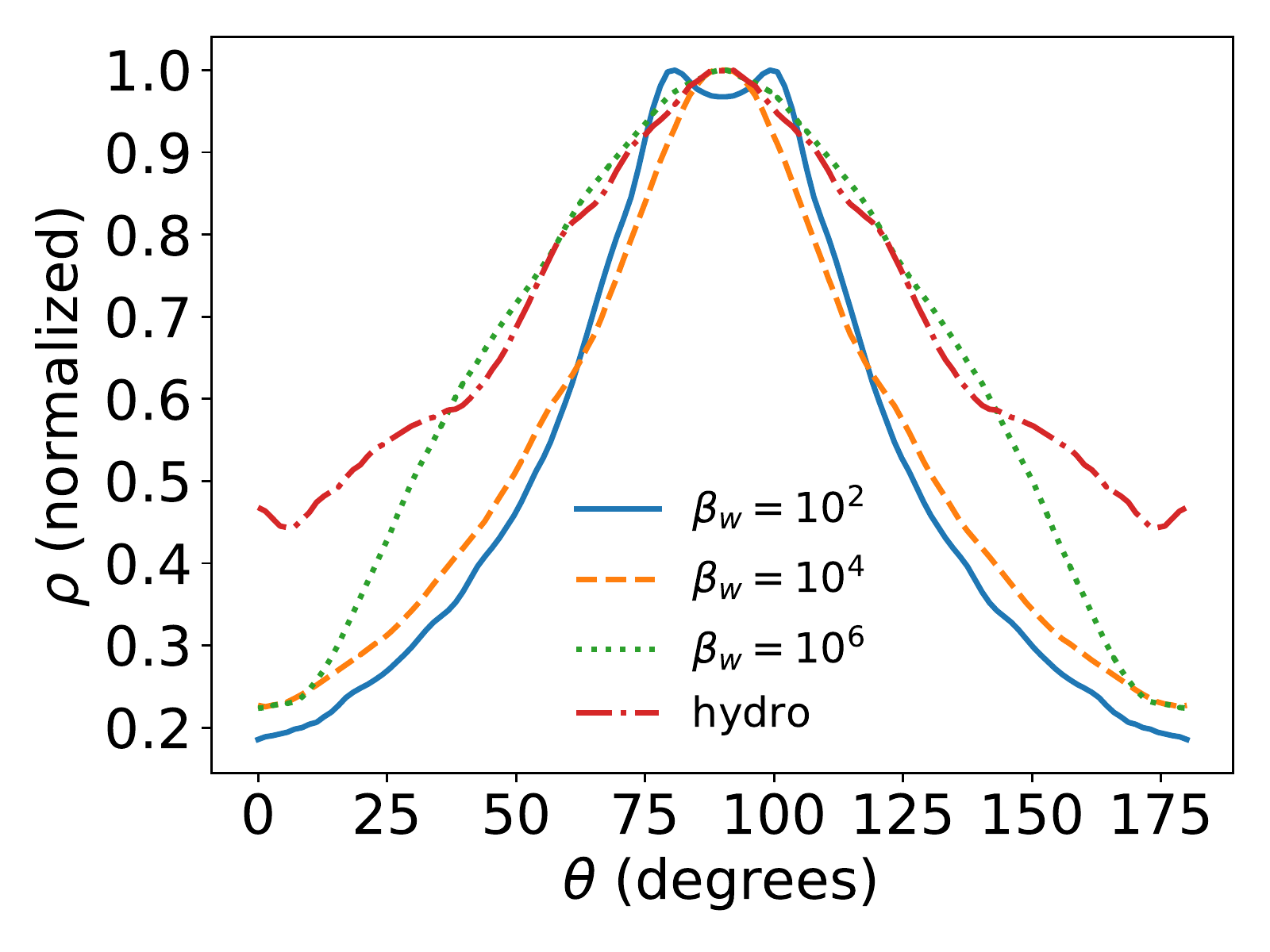}
\caption{Time and $\varphi$ averaged mass density at $r = 5$ mpc, normalized at $\theta=\pi/2$ and folded over the midplane. Strong magnetic fields lead to a larger contrast in density between the midplane and polar regions as compared to the hydrodynamic and more weakly magnetized flows.}
\label{fig:rho_theta}\end{figure}

An additional consequence of the different poloidal dynamics is that the $\beta_w=10^4$ and $10^2$ simulations display a stronger density contrast between the midplane and polar regions compared to the $\beta_w = 10^6$ and hydrodynamic simulations.  This is seen in Figure \ref{fig:rho_theta}, which plots the time and $\varphi$-averaged density folded over the midplane at $r=5$ mpc for the different simulations.  Though the ``disc'' of gas is still quite thick, the equatorial to polar density contrast in the most magnetized case is now a factor of $\sim$ 5 vs. only $\sim$ 2 in the hydrodynamic and more weakly magnetized cases.  This is, however, still a much lower density contrast than typical MHD and GRMHD simulations of MRI driven accretion in tori, which show outflows that are significantly more magnetically dominated, and in which the density at the poles is orders of magnitude less than the midplane.

\begin{figure}
\includegraphics[width=0.45\textwidth]{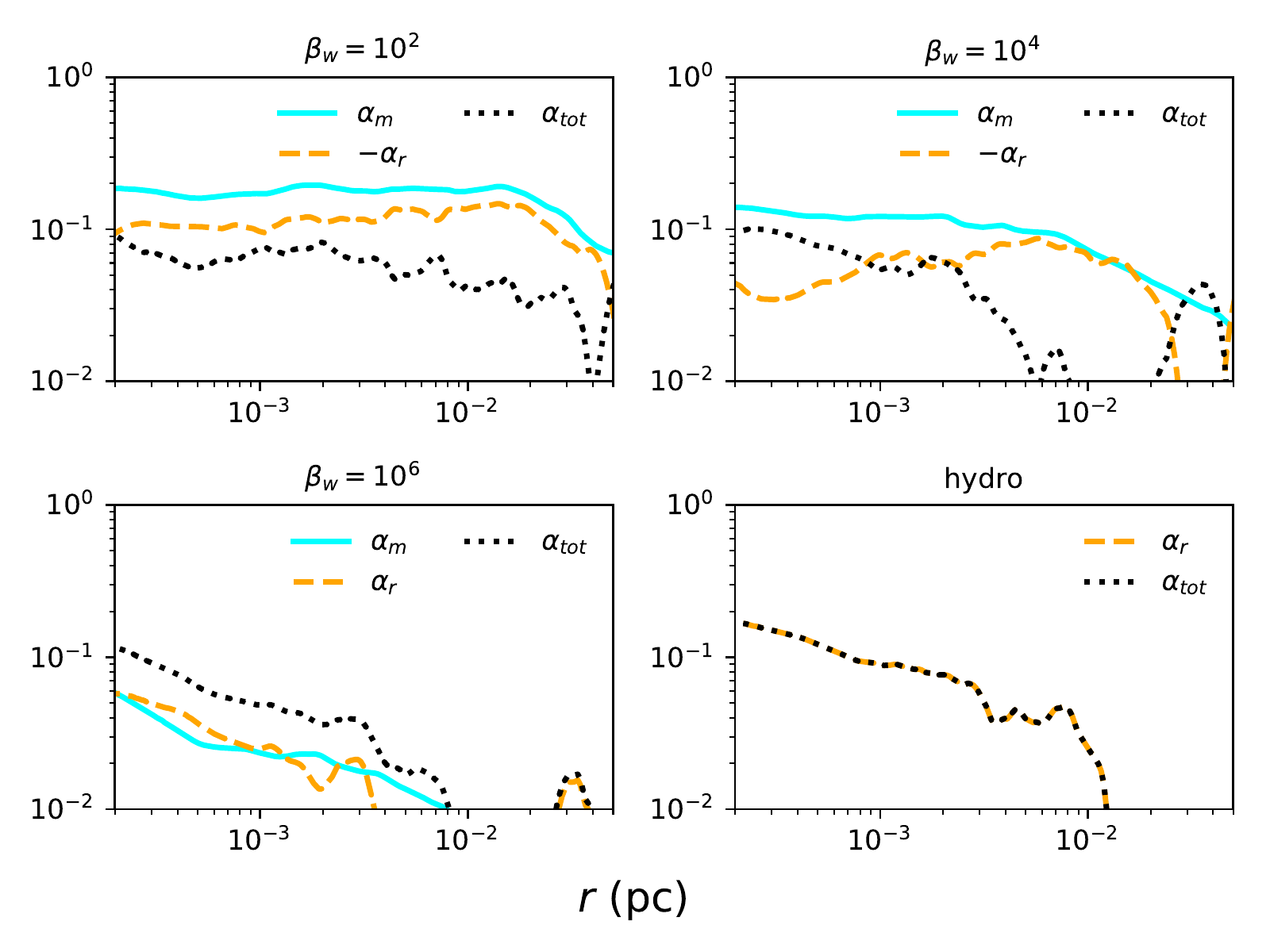}
\caption{Comparison between the time averaged Maxwell (blue solid), Reynolds (dashed yellow), and total (dotted black) $\alpha$ viscosities as defined in \S \ref{sec:stress} for each of our simulations.  In the hydrodynamic case, non-axisymmetric structure and the presence of low angular momentum gas leads to a relatively large ``stress'', and hence, accretion rate. For MHD simulations, while the total stress remains basically unchanged from the hydrodynamic case (see \S \ref{sec:stress}), the Maxwell stress can become larger than the total stress with $\alpha_m \approx 0.2$, with the Reynolds stress becoming negative to compensate.   
}
\label{fig:alpha}
\end{figure}

\subsection{Stresses}
\label{sec:stress}
In order to quantify the relative contribution of the magnetic field to angular momentum transport, we calculate the \citet{Shakura1973} $\alpha$-viscosity from our simulations.  We do this in the same frames defined by the angular momentum direction during the time intervals shown in Figure \ref{fig:L_time_interval} as described in the preceding subsection.

We follow \citet{Jiang2017} by defining the time and angle averaged Reynold's stress
\begin{equation}
S_h \equiv \langle \rho v_r v_\varphi \sin(\theta)\rangle  - \langle \rho v_r  \rangle \langle v_\varphi \sin(\theta)\rangle
\label{eq:Sh}
\end{equation}
and Maxwell stress
\begin{equation}
S_m \equiv \langle B_r B_\varphi \sin (\theta)\rangle.
\label{eq:Sm}
\end{equation}
Then the \citet{Shakura1973} $\alpha$ viscosities are simply $\alpha_h = S_h/P$ and $\alpha_m = S_m/P$, where we have chosen to use the thermal pressure instead of the total (thermal plus magnetic) pressure in the denominator for fair comparison between hydrodynamic and MHD simulations.

The resulting $\alpha$'s for each simulation are plotted in Figure \ref{fig:alpha}.  In the hydrodynamic simulation, the total stress is by definition equal to the Reynolds stress.  This nonzero stress even without magnetic fields or other sources of viscosity can be understood by considering the inflow/outflow structure seen in Figure \ref{fig:mdot_beta_comp}. Accretion occurs via low angular momentum (i.e., low $v_\varphi$) material in the polar regions where the $\varphi$-averaged $v_r$ is large (i.e., close to free-fall) and negative while the midplane consists of high angular momentum (i.e., large $v_\varphi$) material with smaller in magnitude and positive $\varphi$-averaged $v_r$.  Thus, $\langle \rho v_r v_\varphi \sin(\theta)\rangle$ is significantly different than $\langle \rho v_r  \rangle \langle v_\varphi \sin(\theta)\rangle \propto \dot M l $, leading to a large $\alpha_h$.  Thus, $\alpha_h$ is not predominantly a turbulent viscosity but a measure of the fact that there is a superposition of two types of flows: low angular momentum accretion and high angular momentum outflow.   For $\beta_w=10^6$, the Maxwell stress provided by the magnetic field is comparable to Reynolds stress and both work together to transport angular momentum inwards.  This picture is altered for $\beta_w=10^4$ and $\beta_w=10^2$, where the total stress is a competition between a large Maxwell stress and a non-negligible, negative Reynolds stress (where a negative stress implies transport of angular momentum inwards).  For these more magnetized flows, the magnetic field is strong enough to resist being wound up in the $\varphi$ direction, providing significant torque to rotating gas as it falls in. 

In all cases, the total stress, $\alpha_{tot} = \alpha_{h} + \alpha_{m}$, is similar for $r\lesssim 10^{-2}$ pc, varying between $\sim 0.04-0.2$.  This simply reflects the overall dynamical similarity of the flows independent of $\beta_w$. In a steady state accretion flow, the total stress can be written as (from equations \ref{eq:Sh} and \ref{eq:Sm})
\begin{equation}
\alpha_{tot} = \frac{F_J - \langle\dot M \rangle\langle l \rangle }{4 \pi r^3 \langle P\rangle },
\end{equation}
where $F_J = \langle \rho v_r v_\varphi \sin(\theta) \rangle - \langle 4 \pi r^3 B_r B_\varphi \sin(\theta) \rangle =$ is the constant flux of angular momentum 
and $l$ is the specific angular momentum.  The constant $F_J$ is set by the accretion rate and angular momentum at the inner boundary and is generally small.  Thus, since $\dot M$ and $l$ are relatively unchanged in an angle averaged sense going from hydrodynamics to MHD, the total stress is unchanged.  

\subsection{MRI}
\label{sec:MRI}
We have shown (Figure \ref{fig:beta_comp}) that the magnitude of the magnetic field at small radii is only weakly dependent on $\beta_w$, the parameter governing the strength of the magnetic field in the stellar winds.  A natural mechanism to explain this is the magnetorotational instability, which can amplify an arbitrarily small magnetic field to reach $\beta \lesssim 10$ in differentially rotating flows, such as we have here. However, we have also shown that the gas in our simulations never circularizes and therefore does not spend many orbits at small radii.  Consequently, there is not sufficient time in a Lagrangian sense for the MRI to grow.  

We can further evaluate the role of the MRI by using an estimate of the fastest growing wavelength for perturbations given by
\begin{equation}
\lambda_{MRI,\theta} \approx \frac{2 \pi |B_\theta|}{\sqrt{4 \pi \rho}\Omega},
\end{equation}
where $\Omega \equiv v_\varphi/(r \sin(\theta))$ is the rotational frequency.  At least two criteria need to be met in order for the MRI to operate in numerical simulations: 1) $\lambda_{MRI,\theta}$ needs to be resolved, that is, the cell length $\Delta x$ needs to be $\ll$ $\lambda_{MRI,\theta}$, and 2) $\lambda_{MRI,\theta}$ needs to be smaller than the scale height of the disc, otherwise the perturbations have no room to grow.  

\begin{figure}
\includegraphics[width=0.45\textwidth]{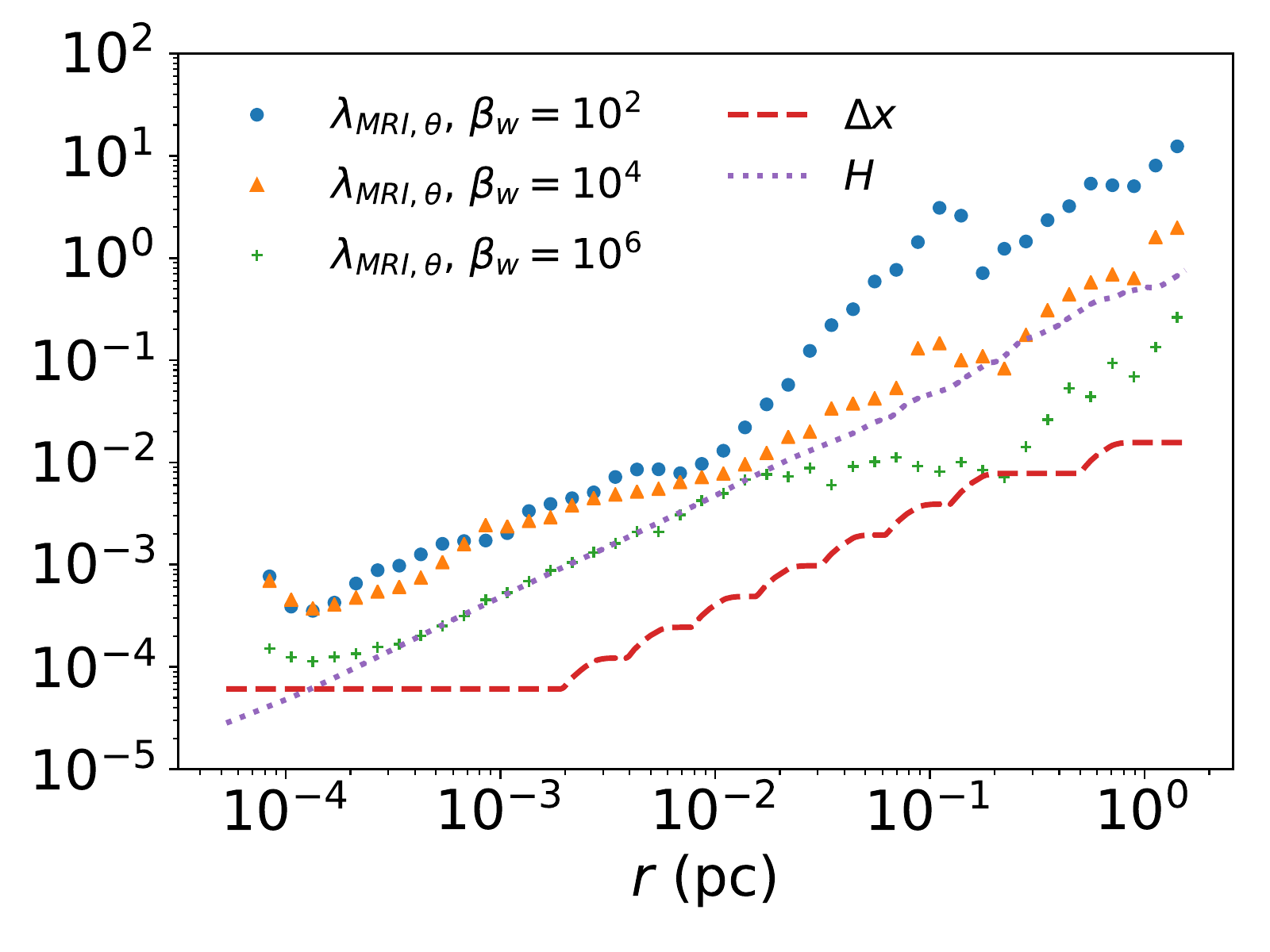}
\caption{Wavelength of the most unstable mode for the MRI computed in the midplane for our three MHD simulations as compared to the scale height of the disc, $H$, and the resolution of our grid, $\Delta x$.  $\lambda_{MRI,\theta}$ is well resolved but is larger than the scale height of the disc.  The MRI is suppressed by the strong $\beta \sim$ few magnetic field (Figure \ref{fig:beta_comp}) produced by compression as the gas flows in.  }
\label{fig:lambda_mri}
\end{figure}

Figure \ref{fig:lambda_mri} shows $\lambda_{MRI,\theta}$ in the midplane of the disc compared to the scale height of the disc, defined as $H \equiv r \langle \rho |\theta-\pi/2| \rangle/\langle \rho \rangle $, and the resolution of our grid.  We find that $\lambda_{MRI,\theta}$ is sufficiently resolved at all radii but that it is larger than the scale height for all of our MHD simulations.  This implies that even if the gas in our simulations were to circularize, which we reiterate does not in fact occur, the MRI would have no room to operate.  Therefore, we conclude that the MRI is not an important source of magnetic field amplification or angular momentum transport in our simulations. This finding, along with the fact that $\beta$ $\sim$ a few at small radii in our lower $\beta_w$ runs, is also observed in MAD simulations \citep{Igumenshchev2003,White2019}.   Instead of the MRI, we explain the saturation of the magnetic field at small radii displayed  in Figure \ref{fig:beta_comp} with simple compression/flux freezing.  An initially weak field at large radii will be compressed as it is pulled inwards by the bulk motion of the gas.  It will continue to do so until $\beta\sim$ a few, when the field becomes dynamically important and starts to resist the fluid motion.  At this point, the field maintains $\beta\sim$ a few as it continues to accrete. For small $\beta_w$, this happens at large radii, while as $\beta_w$ increases the field reaches $\beta$ $\sim$ a few at progressively smaller radii.  If we were able to reach even smaller radii with our simulations, we predict that even the $\beta_w =10^6$ run will ultimately reach $\beta$ of $\sim$ a few and the field would become dynamically important (see also \S \ref{sec:extrapolate}).

\subsection{Magnetic Field Structure}
\label{sec:field_structure}

\begin{figure}
\includegraphics[width=0.49\textwidth]{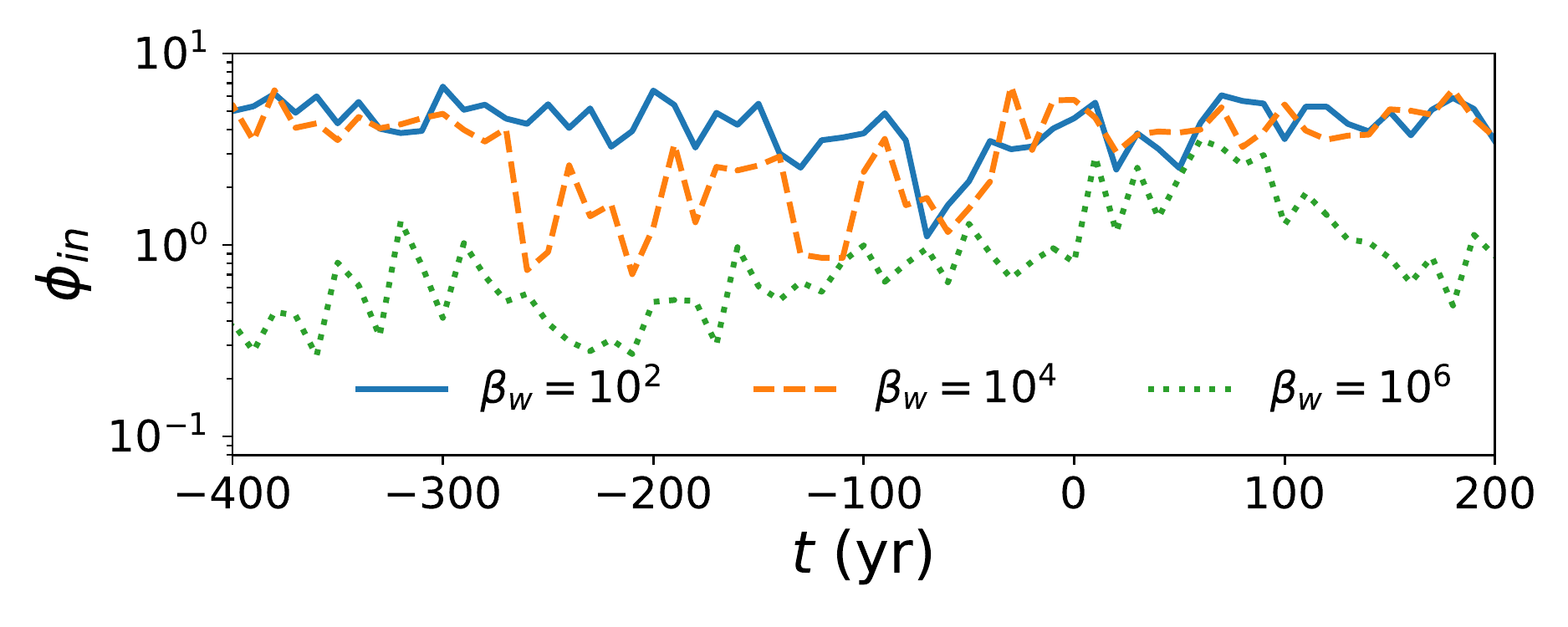}
\caption{Magnetic flux threading the inner boundary as a function of time for our three MHD simulations, $\phi_{in}$, in units such that the MAD value is $\sim 50$ in GR (see Equation \ref{eq:phi_in}). Orders of magnitude difference $\beta_w$ corresponds to only a factor of $\sim$ few difference in $\phi_{in}$ because the field strength at small radii is only weakly sensitive to $\beta_w$ (Figure \ref{fig:beta_comp}). In all cases $\phi_{in}$ is $\lesssim 10 \%$ of the GR MAD limit, but the $\beta_w=10^2$ simulation is near or has reached the expected Newtonian MAD limit of $\lesssim 10$ appropriate for these simulations.  
 }
\label{fig:phibh_comp}
\end{figure}

We now turn our attention to the structure of the magnetic field at small radii.  RQS19 predicted that the amount of magnetic flux ultimately threading the inner radii of the domain, $\phi_{in}$, is roughly insensitive to $\beta_w$ and roughly constant in time, falling between $\approx 1-6$ in units such that the Magnetically Arrested Disc (MAD) limit in GR is $\approx$ 50 \citep{Sasha2011}. To demonstrate this result more clearly, Figure \ref{fig:phibh_comp} plots $\phi_{in}$, defined as
\begin{equation}
\phi_{in} \equiv \left. \frac{1/2\int |B_r| r^2 d\Omega }{ r\sqrt{|\dot M | v_{kep}}}\right|_{r = r_{in}},
\label{eq:phi_in}
\end{equation}
for our three MHD simulations. Across four orders of magnitude in $\beta_w$, $\phi_{in}$ varies by only a few, and for each run it is roughly constant in time.  Averaged over the interval (-100 yr, 100 yr), the values are 4.4, 3.5, and 1.1 for $\beta_w$ of $10^2$, $10^4$, and $10^6$, respectively.
These differences in $\phi_{in}$ are even smaller when extrapolated to smaller radii, which we do in \S \ref{sec:extrapolate}.  Briefly, we expect that for all reasonable $\beta_w$, $\phi_{in}$ at the horizon will be around the $\beta_w=10^2$ value shown in Figure \ref{fig:phibh_comp}, independent of $\beta_w$.  The result that $\phi_{in}$ becomes quasi-steady despite the fact that matter is continually being accreted (bottom panel of Figure \ref{fig:mhd_comp}) is noteworthy.   We hypothesize that this is a consequence of the magnetic field being accreted changing direction with time so that the incoming field reconnects with the field in the boundary in a way that regulates the value of $\phi_{in}$. If instead the incoming field had the same orientation at all times, $\phi_{in}$ would show a continual rise until the field threading the boundary became strong enough to arrest accretion.   Alternative possibilities include that the outflow preferentially removes magnetic fields, or that a balance of advection and diffusion regulates the value of $\phi_{in}$ (as seen in simulations of magnetically ``elevated'' discs, \citealt{Zhu2018,Mishra2019}).

It is important to note, however, that the amount of net magnetic flux threading the event horizon required for a simulation to reach the MAD state in GR ($
\phi_{in}$ $\approx$ 50) is not necessarily the same for the Newtonian simulations we have here.  GR effects cause gravity near the event horizon to be effectively stronger, requiring more magnetic flux (i.e. more magnetic pressure) to arrest the accretion flow.  In Newtonian simulations the threshold value for $\phi_{in}$ is likely lower.   To effectively arrest the flow, the magnetic field must be strong enough to exert an outward radial force that is at least as large the radial ram pressure, $\rho v_r^2$, and perhaps as large as the gravitational force.  Conservatively, then, if we assume that a MAD state is reached when the magnetic pressure at the inner boundary roughly balances gravity, then by equating the gradient of the magnetic pressure with $\rho G M/r^2$ one finds a rough threshold value of $\phi_{in}$ $\sim$ $ 2 \pi \left. \sqrt  {v_{kep}/v_{r}}\right|_{r=r_{in}}$, on the order of $6-10$ if $v_r$ is a little less than free fall at $r=r_{in}$ as it is here.  Note that in deriving this threshold value on $\phi_{in}$ we have assumed that $B^2$ scales roughly as $r^{-2}$ (Figure \ref{fig:beta_comp}) and neglected the contribution of magnetic tension.  Relaxing these assumptions, the dashed line in Figure \ref{fig:press_radius} plots the outward Lorentz force (including both magnetic pressure and magnetic tension forces, $\propto$ $[\{\mathbf{\nabla} \times \mathbf{B}\} \times \mathbf{B}] \cdot \hat r$) relative to the gravitational force.  We find that the Lorentz force provided by the magnetic field is a factor of 10 smaller than the gravitational force at the inner boundary even though $\phi_{in}$ is near the previous simple estimate of the MAD threshold.  On the other hand, Figure \ref{fig:press_radius} also shows that the Lorentz force is as large as or larger than the $v_r$ ram pressure.  So while our simulations do not appear to be fully magnetically arrested based on the fact that the accretion rate is comparable in both hydrodynamic and MHD simulations (Figure \ref{fig:mhd_comp}), the amount of magnetic flux at the inner boundary must be only modestly less than the MAD limit.

One possible concern is that just outside of the inner boundary our simulations are unresolved, where $r/\Delta x  \gtrsim 2$ and $\Delta x$ is the size of an edge of a cubic cell.  This could potentially lead to numerical diffusion of magnetic fields and prevent a larger amount of flux from accumulating.  Two things suggest that this locally limited resolution is not effecting our calculation of $\phi_{in}$.  First, $\phi(r)$, i.e., Equation \eqref{eq:phi_in} evaluated at a radius $r$ instead of $r_{in}$, is roughly independent of $r$ in the $\beta_w=10^2$ simulation, equal to $\phi_{in}$. This includes larger radii that are much better resolved where the typical grid spacing is $r/\Delta x  \sim 128$. Second (and more convincingly), we ran an additional  $\beta_w=10^2$ simulation with an inner boundary radius that was a factor of 8 times the size of the smallest cell edge instead of our fiducial factor of 2, meaning that the region just outside the boundary was 4 times as well resolved.  This additional simulation displayed approximately the same value of $\phi_{in}$ as a simulation with the same $r_{in}$ but only 2 cells per $r_{in}$.  Ultimately, much better resolved simulations will be required to definitively assess the impact of numerical diffusion on the values of $\phi_{in}$ determined here.

\begin{figure}
\includegraphics[width=0.45\textwidth]{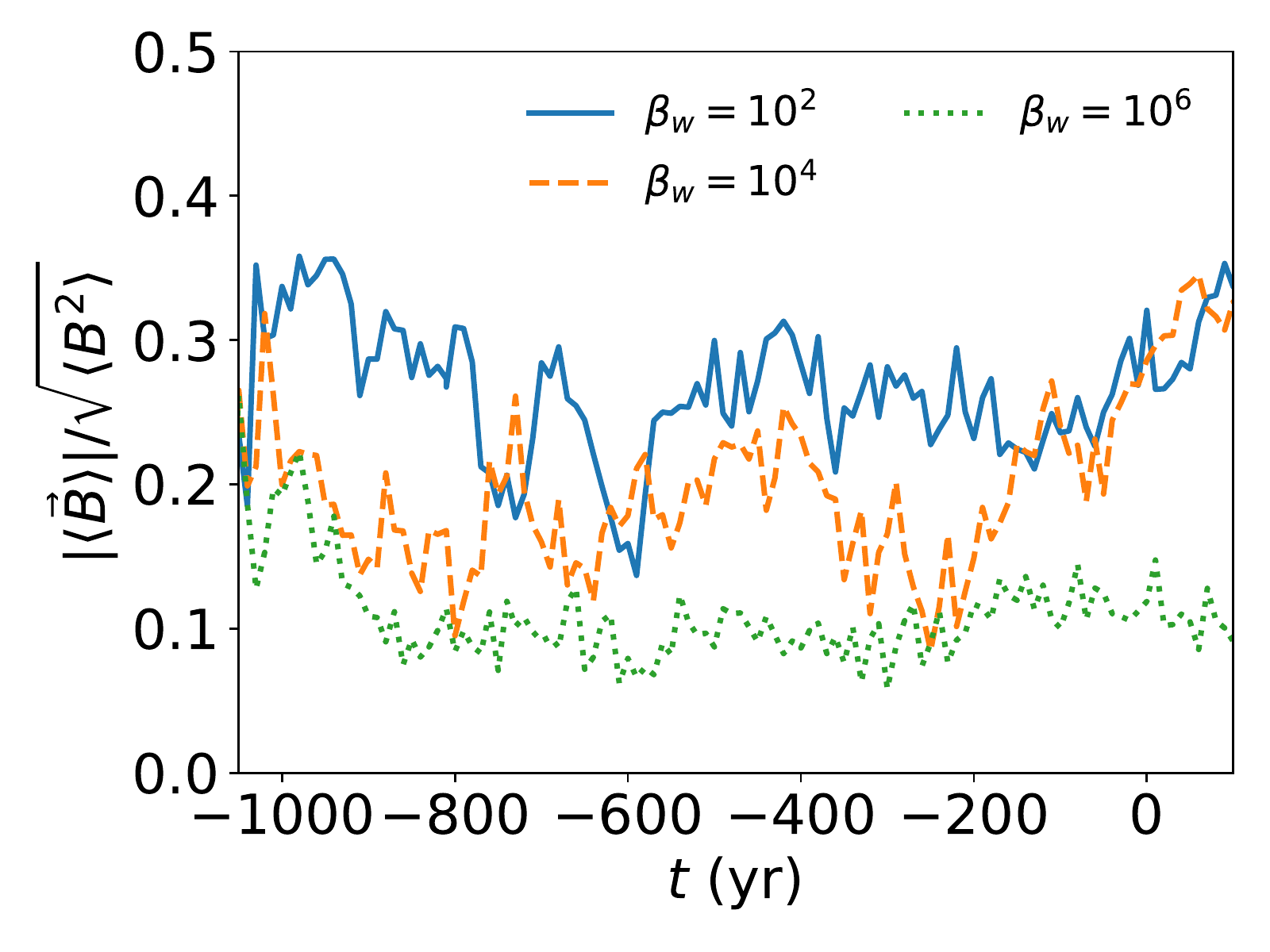}
\caption{Magnitude of the time and angle-averaged magnetic field vector divided by the rms magnitude of the field for $\beta_w=10^2$ (solid blue), $\beta_w=10^4$ (dashed orange), and $\beta_w=10^6$ (dotted green).  This measures the degree to which the magnetic field is ordered, and increases with decreasing $\beta_w$ because stronger fields are able to resist fluid motion and more effectively retain a coherent structure.   }
\label{fig:B_order_comp}\end{figure}

\begin{figure}
\includegraphics[width=0.45\textwidth]{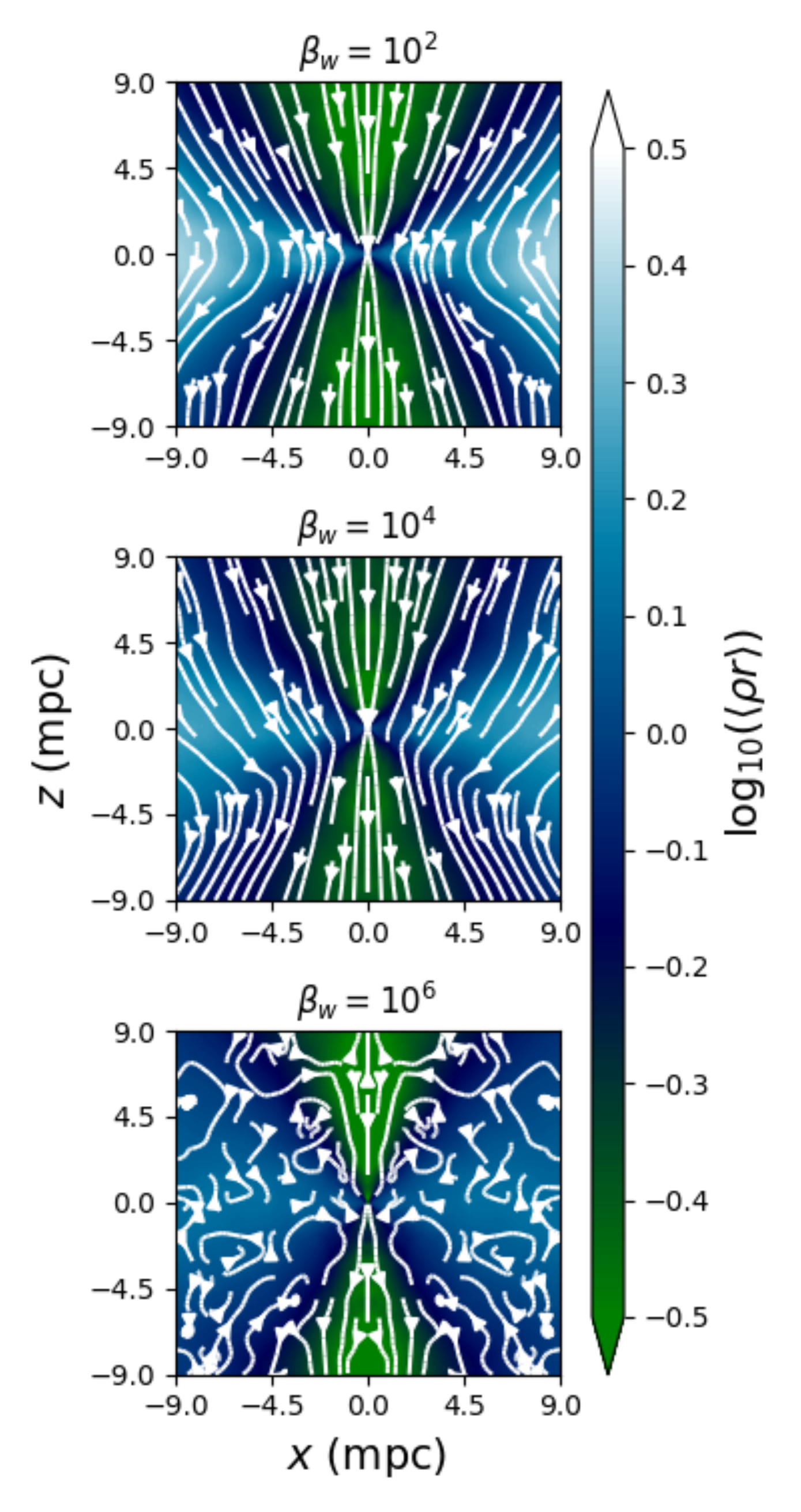}
\caption{Time and $\varphi$-averaged mass density weighted by radius in a coordinate system such that the $z$-direction is aligned with the angular momentum of the gas (Figure \ref{fig:L_time_interval}), normalized, and overplotted with magnetic field lines for $\beta_w=10^2$ (top), $10^4$ (middle), and $10^6$ (bottom).  The stronger the field, the more it is able to resist being dragged along by the random motions of the flow and retain a coherent structure   }
\label{fig:fieldlines_comp}\end{figure}

We quantify the relative strength of the vertical magnetic field by computing the ratio between the magnitude of the average magnetic field vector, $|\mathbf{B}|$, to the root-mean-squared field strength, $\sqrt{\langle B^2\rangle}$. For a completely vertical field this quantity would be 1, while for a completely toroidal or random field it would be 0.  Figure \ref{fig:B_order_comp} plots $\langle |\mathbf{B}| \rangle /\sqrt{\langle B^2 \rangle}$ averaged over angle and the inner $5 \times 10^{-4}$ pc to $3 \times 10^{-2}$ pc in radius for $\beta_w = 10^2,10^4,$ and $10^6$, where we find that the relative strength of the ordered field increases with decreasing $\beta_w$.  This same trend is seen in the poloidal field lines (Figure \ref{fig:fieldlines_comp} where they are plotted on top of mass density), where the direction of the field goes from mostly random at $\beta_w=10^6$, to nearly vertical at $\beta_w=10^2$.  The weaker the magnetic field, the more it is able to be twisted by the motion of the gas and lose its original structure.  

The quantities plotted in Figures \ref{fig:B_order_comp} and \ref{fig:fieldlines_comp} do not effectively probe the $\varphi$ component of the field, which in principle could be significant.  To quantify this, we compare the relative strength of the mean $B_\varphi$ to the mean $B_r$ and $B_\theta$ field components. We define an `antisymmetric' average of $B_r$ as 
\begin{equation}
\langle \tilde B_r \rangle = \int\limits_{t_1}^{t_2}\int\limits_0^{2\pi} \int\limits_{0}^{\pi/2}  B_r d\theta d\varphi dt - \int\limits_{t_1}^{t_2}\int\limits_0^{2\pi}\int\limits_{\pi/2}^\pi B_r d\theta d\varphi dt ,
\label{eq:Br_antisym}
\end{equation}
where $t_1$ and $t_2$ are the endpoints of the time interval for averaging.  The minus sign in Equation \eqref{eq:Br_antisym} prevents the radial field from averaging to zero over all angles. 
For $\beta_w=10^6$, the toroidal field dominates with $\langle B_\varphi \rangle^2/\left(\langle \tilde B_r\rangle^2 + \langle B_\theta \rangle^2 + \langle B_\varphi \rangle^2\right) \approx 0.8-1$ for $r \lesssim 2 \times 10^{-2}$ pc because the field is weak enough to be completely stretched out by the orbital motion of gas.  For $\beta_w=10^4$, on the other hand, the field is able to resist the orbital motion (seen also in the torque that it exerts; Figure \ref{fig:alpha}) and retain a predominantly poloidal structure, with $\langle B_\varphi \rangle^2/\left(\langle \tilde B_r\rangle^2 + \langle B_\theta \rangle^2 + \langle B_\varphi \rangle^2\right)$  $ \lesssim 0.2$  for $r\lesssim 3 \times 10^{-3}$ pc. This is even more true for $\beta_w=10^2$, which has $\langle B_\varphi \rangle^2/\left(\langle \tilde B_r\rangle^2 + \langle B_\theta \rangle^2 + \langle B_\varphi \rangle^2\right) \lesssim 0.1$  for $r\lesssim 4 \times 10^{-3}$ pc.

\begin{figure}
\includegraphics[width=0.45\textwidth]{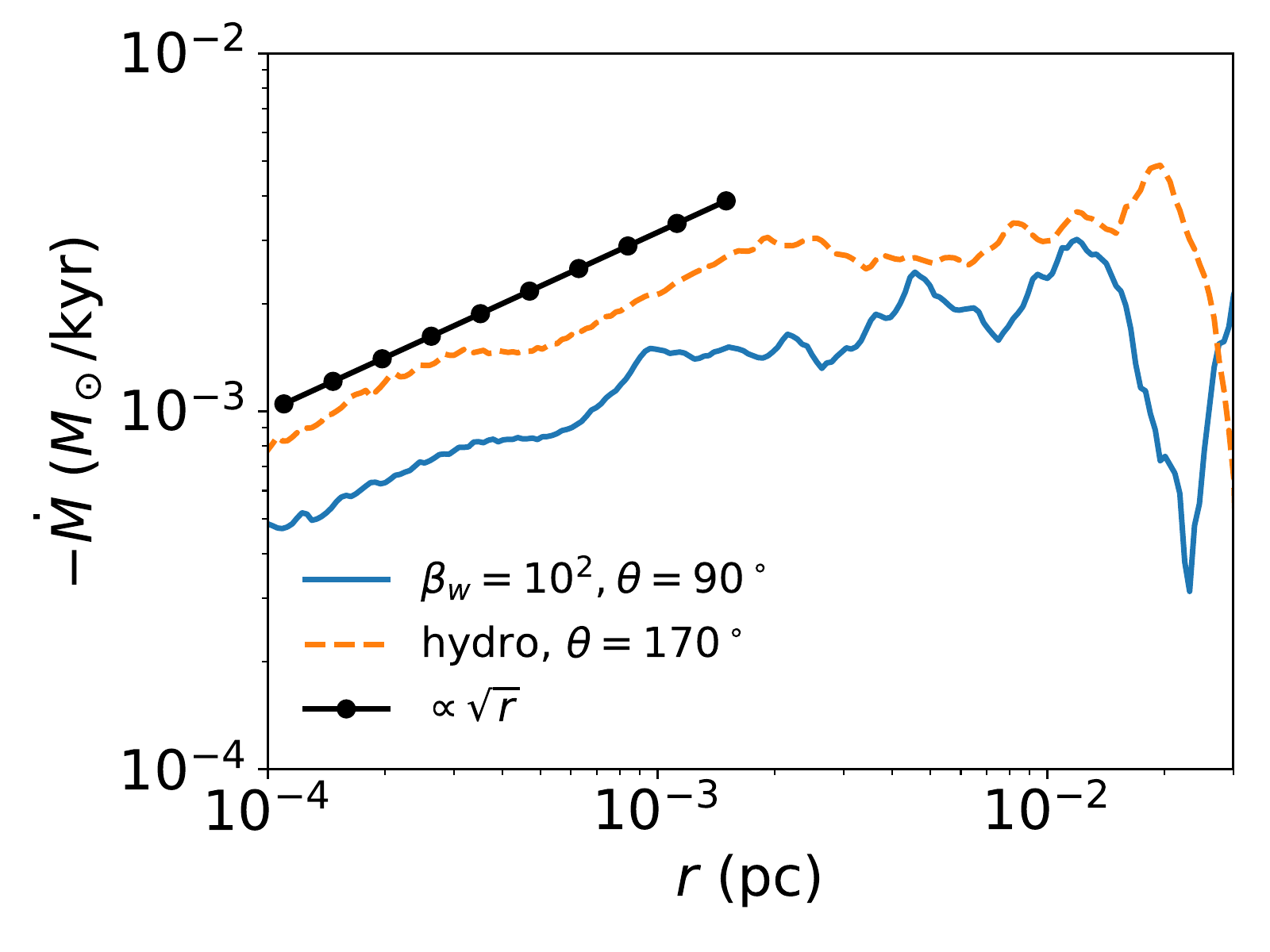}
\includegraphics[width=0.45\textwidth]
{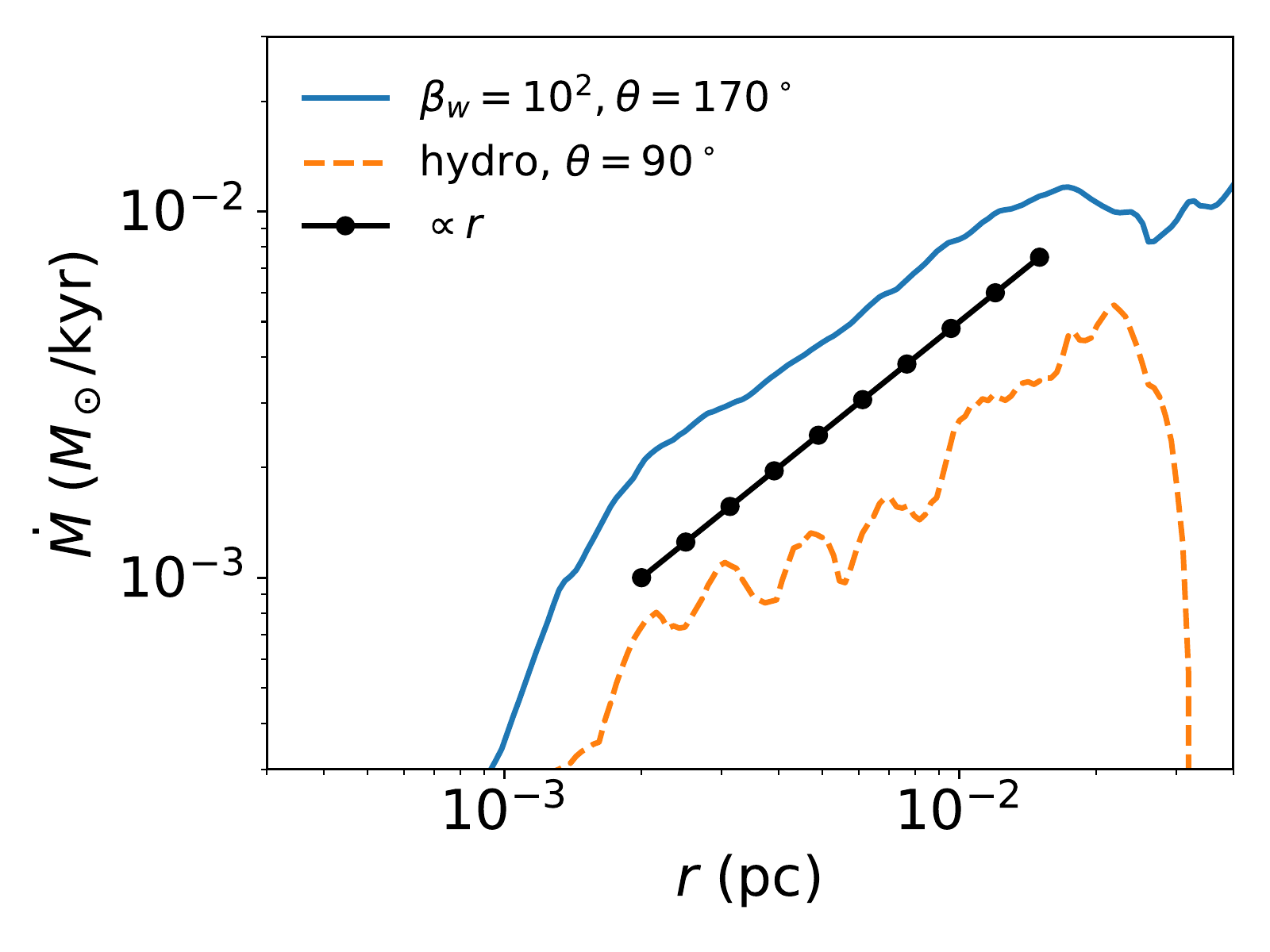}
\caption{Top: Comparison between the time and $\varphi$-averaged accretion rate in the midplane for the $\beta_w=10^2$ simulation (solid blue) vs. the pole for the hydrodynamic simulation (dashed orange).  Both of these regions are dominated by inflow and show $\dot M \propto \sqrt{r}$ as expected for the $d \dot M/dl \approx \textrm{const.}$ distribution provided by the stellar winds. Bottom: Comparison between the time and $\varphi$-averaged accretion rate in the pole for the $\beta_w=10^2$ simulation (solid blue) vs. the midplane for the hydrodynamic simulation (dashed orange).  Both of these regions have a net outflow and show $\dot M \propto r $, implying a roughly constant velocity outflow since $\rho \propto r^{-1}$ (Figure \ref{fig:mhd_comp}).  }
\label{fig:mdot_in_out_comp}\end{figure}

\subsection{Physical Interpretation of The Role of Magnetic Fields}
\label{sec:interp}
Thus far we have presented seemingly paradoxical results.  On one hand, for sufficiently magnetized stellar winds (e.g., $\beta_w=10^2,10^4$), the magnetic field at small radii reaches near equi-partition with the plasma, achieving $\beta\sim$ a few, reversing the polar inflow seen in hydrodynamic simulations, and driving accretion in the midplane.  On the other hand, the net accretion rate through the inner boundary and the radially averaged fluid quantities are largely unaffected by the presence of magnetic fields.  How can this be?  In the conventional picture of MRI driven accretion from a rotationally supported torus, it would require an improbable cooincidence, where the midplane accretion driven by the MRI exactly equals the original hydrodynamic polar accretion despite the fact that they are governed by different physical considerations.  As we have shown, however, our simulations do not fit this conventional picture.  The gas with significant angular momentum clearly does \emph{not} circularize into a configuration where the velocity is primarily in the azimuthal direction (e.g., Figures  \ref{fig:time_series_hydro} and \ref{fig:time_series_beta_1e2}), but instead retains significant radial velocity of order free fall throughout the domain.  Simply put, gas accreting from large radii is quick to either flow through the inner boundary or flow right back out.  Magnetic fields are not strong enough to modify these flows by more than order unity even at $\beta \sim$ 1.  Moreover, even in the hydrodynamic simulation, inflow is not occurring only in the poles as Figure \ref{fig:mdot_beta_comp} would imply but at all polar angles.  It is only in an azimuthally-averaged sense that $v_r$ is positive and small in the midplane because there is also significant outflow present (at different $\varphi$).  The primary role of magnetic fields, then, is not to drive accretion but to redirect the outflow from the midplane to the pole.  This means that 1) the same physical processes govern accretion in the hydrodynamic and MHD simulations and 2) the net accretion rate is essentially determined by hydrodynamic considerations, namely, the distribution of angular momentum at large radii, a quantity set by the winds of the WR stars. 

The lack of circularization in our simulations, the crucial factor in determining this accretion structure, is at least in part due to radiative cooling being inefficient at removing dissipated energy in the gas streamers seen in Figures \ref{fig:time_series_hydro} and \ref{fig:time_series_beta_1e2}. As the gas comes in along nearly parabolic orbits it heats up and (because it can't cool) expands outward, making it more difficult for it to circularize. This is analogous to the difficulty that simulations of tidal disruption events have in forming a circular accretion disc (for recent discussion, see, e.g., \citealt{Stone2019,Lu2019}).

\subsection{Dependence On The Inner Boundary Radius}
\label{sec:extrapolate}
The simulations we have performed, while modeling a radial range of just over 3 orders of magnitude, are not able to penetrate all the way to the event horizon of Sgr A* but have inner boundary radii still a few hundred times farther out.  Thus, it is important for us to understand how the artificially large inner boundary of our simulation (which acts as the black hole) affects the results. By varying the inner boundary, RQS18 showed that the predicted accretion rate through the inner boundary in our hydrodynamic simulations is $\dot M \approx 2.4 \times 10^{-8}$ $(r_{in}/r_G)^{1/2}$ $M_{\odot}$/yr, where the dependence on $r_{in}$ is set by the distribution of accretion rate with angular momentum at large radii; for a smaller inner boundary radius, less material has angular momentum low enough to ultimately accrete. This predicted accretion rate is consistent with both observational constraints and emission models (see RQS18).  

In MHD, we have shown that even for strong magnetic fields the radially averaged fluid variables are mostly unchanged going from hydrodynamics to MHD (Figure \ref{fig:mhd_comp}), including the accretion rate. Thus the above relation between $\dot M$ and $r_{in}$ still holds.
As we have argued in the preceding section, this counter-intuitive result is a consequence of the fact that the supply of infalling gas at small radii is still mostly set by the distribution of accretion rate with angular momentum at large radii.  The gas provided by nearby stellar winds has a typical distribution of $d \dot M/ d l \approx$ const. which results in $\dot M_{in} \propto$ $\sqrt{r}$ (see Appendix A in RQS18).  Note that since the winds emit at all angles, this is the distribution of infalling gas for both the poles and the midplane. Figure \ref{fig:mdot_in_out_comp} confirms this expectation, showing that $\dot M \propto \sqrt{r}$ for the inflow in both hydrodynamics (in the polar region) and $\beta_w=10^2$ MHD (in the midplane).  

Figure \ref{fig:beta_comp} shows that $\beta$ tends to decrease with decreasing radius until it reaches $\sim$ a few, at which point it becomes independent of radius. For $\beta_w=10^2$ $\beta$ is $\approx$ 1-2 and roughly constant throughout the domain, for $\beta_w=10^4$ it decreases from $\beta \approx 200$ at large radii to $\beta \approx$ 2 at $r \approx 6 \times 10^{-4}$ pc and remains constant for $r \lesssim 6 \times 10^{-4}$ pc, while for $\beta_w=10^6$ it decreases from $\beta \approx 2\times 10^4$ to $\beta \approx$ 4 near the inner boundary.  It is natural to suppose that if the inner boundary radius of the simulation was reduced then the $\beta_w=10^6$ run would ultimately also reach $\beta \approx 2$. Regrettably this is not something we can test with our current computational resources; however, we can \emph{increase} the size of the inner boundary and infer how $\beta$ depends on $r_{in}$ in the same way that we used to extrapolate $\dot M$ in RQS18. Doing so we estimate that all models with $\beta_w \le 10^7$
will reach $\beta$ of $\sim$ a few by 2 $r_g$ (the event horizon radius of a non-rotating black hole).  Thus $\beta_w \sim 10^7$ is a critical value that determines whether or not the horizon scale accretion flow will more closely resemble the hydrodynamic simulations ($\beta_w \gtrsim10^7$) or the more magnetized wind simulations ($\beta_w \lesssim10^7$).

\begin{figure}
\includegraphics[width=0.45\textwidth]{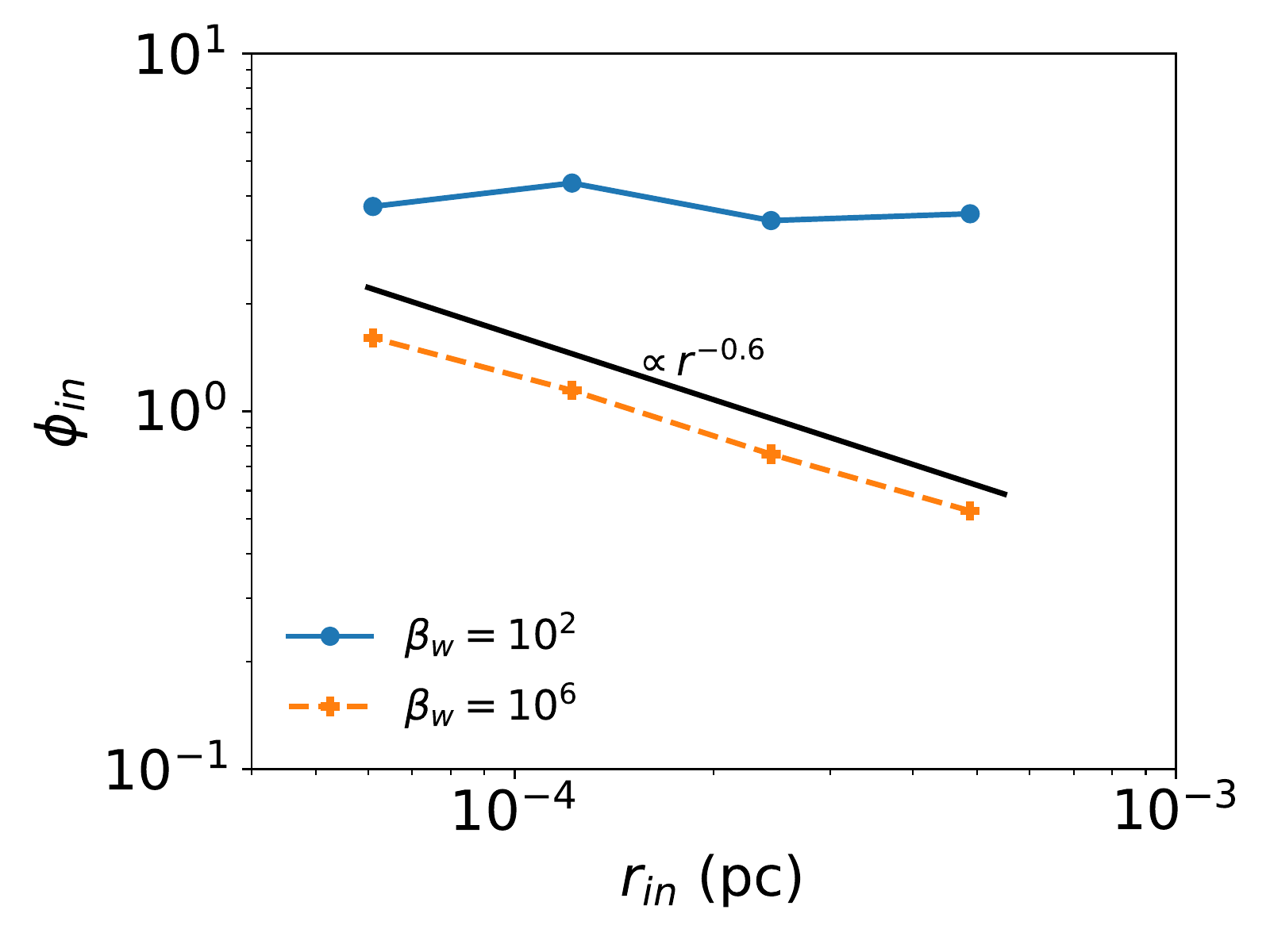}
\caption{Magnetic flux threading the inner boundary of our simulation, $\phi_{in}$ (Equation \ref{eq:phi_in}), averaged over the time interval (-100 yr,100 yr) and plotted as a function of inner boundary radius for $\beta_w=10^2$ (blue solid) and $\beta_w=10^6$ (orange dashed). $\phi_{in}$ is independent of $r_{in}$ for $\beta_w=10^2$ where $\beta$ has reached $\sim$ few (Figure \ref{fig:beta_comp}), while it increases with decreasing $r_{in}$ for $\beta_w=10^6$ because $\beta$ is decreasing with decreasing $r_{in}$.  For $\beta_w=10^6$ we find $\phi_{in}$ $\tilde \propto$ $r^{0.6}$ and thus expect it to reach $\sim$ 5 (the $\beta_w=10^2$ value) for an inner boundary at the event horizon. }
\label{fig:phi_in_rin}\end{figure}

Similar behavior is seen with the magnetic flux threading the inner boundary.  Figure \ref{fig:phi_in_rin} shows the time-averaged $\phi_{in}$ for $\beta_w=10^2$ and $\beta_w=10^6$ and  four values of the inner boundary radius.  As was the case for $\beta$, $\phi_{in}$ is independent of $r_{in}$ for $\beta_w=10^2$.  This is again because the $\beta_w=10^2$ simulation has already reached $\beta \sim $ few at $r\gg r_{in}$.  Since $\dot M \propto r_{in}^{1/2}$ $v_{kep}(r_{in})\propto r_{in}^{-1/2}$, Equation \eqref{eq:phi_in} gives $\phi_{in}$ $\approx$ const. For $\beta_w=10^6$ on the other hand, $\phi_{in}$ increases with decreasing $r_{in}$.
Empirically, we find in Figure \ref{fig:phi_in_rin} that for $\beta_w=10^6$, $\phi_{in} \propto r_{in}^{0.6}$, predicting that it will reach $\approx$ 5 by $r_{in} = 3 \times 10^{-6}$ pc $\approx$ 20 $r_g$.   At that point, we expect $\phi_{in}$ to stop increasing in the same way that $\phi_{in}$ is independent of $r_{in}$ for $\beta_w=10^2$.

\section{Comparison to Previous Work}
In analyzing our simulations, we have found it instructive to compare and contrast our results with previous simulations in the literature that considered the problem of accretion onto Sgr A* and related systems via large scale feeding. In this section we do so for two key works.
\subsection{Proga \& Begelman 2003}
\label{sec:proga}
\citet{PB03A,PB03B} (hereafter PB03A and PB03B) presented the results of 2D inviscid hydrodynamic (PB03A) and MHD (PB03B) simulations of accretion onto supermassive black holes as fed by gas with a $\theta$-dependent distribution of angular momentum at large radii.  This approach differs from the standard method of initializing simulations with equilibrium tori without any feeding at large radii and is perhaps a better approximation of the feeding of gas via stellar winds in the Galactic Centre.  In fact, in many ways the results of PB03A and PB03B are strikingly similar to the results of RQS18 and those presented here.  We both find that accretion in hydrodynamic simulations occurs via low angular momentum gas falling in mostly along the polar regions while the higher angular momentum midplane is (on average) outflowing.  We both also find that this structure is reversed in MHD for sufficiently large magnetic fields, with the low angular momentum polar inflow getting quenched by magnetically driven polar outflow while gas in the midplane accretes.   PB03B, however, found that the accretion rate in the MHD case was significantly reduced compared to the hydrodynamic case because the induced midplane accretion was not enough to compensate for the loss of polar inflow.  In this work, on the other hand, the midplane accretion in MHD seems to roughly equal the original hydrodynamic polar inflow so that the net accretion rate is relatively the same in MHD and hydrodynamics. 

The key difference lies in the structure of the high angular momentum gas in the midplane.  PB03A found that this gas was able to circularise and build up into a nearly constant angular momentum torus that blocked the inward flow of gas for polar angles close to the equator.  We find that the high angular momentum gas in our hydrodynamic simulation never circularises but mostly flows right back out after falling in to small radii.  This is more easily accomplished in 3D where flow streams can avoid intersecting;  in 2D axisymmetry (used in PB03A and PB03B), collisions between the infalling and outflowing high angular momentum gas are unavoidable and can dissipate radial kinetic energy and efficiently circularise the material. Because of this circularisation in PB03A, by adding even a weak magnetic field, the MRI is able to grow as the gas in the torus orbits and becomes the dominant driver of accretion. Thus, the accretion rate in PB03B is mostly set by completely different physical considerations (the properties of the MRI) than in PB03A (the availability of low/zero angular momentum gas).   In our simulations, however, even in MHD the dominant source of accretion is still the supply of low angular momentum gas with an order unity correction for global torques provided by strong magnetic fields that have been compressed to $\beta$ $\sim$ few at small radii.  This means that the local supply of mass available to accrete is set mostly by hydrodynamic considerations (i.e., the distribution of angular momentum vs. accretion rate provided by the nearest stellar winds).

One of the main conclusions of PB03B was that the MRI driven accretion seen in their simulations was roughly independent of the angular momentum distribution of material sourced at large radii, ultimately resembling simulations that are initialized with an equilibrium torus seeded by a weak magnetic field. This served as partial motivation and justification for future work to mostly ignore large radii and instead focus on horizon scale ($\lesssim 100-1000 r_g$) simulations starting from equilibrium tori.  Our results, however, suggest that when a more complicated treatment of accretion sourced by stellar winds in full 3D is considered, the properties of the accretion flow at small radii are very strongly influenced.  We suspect that the major source of this difference is the non-axisymmetric nature of how the gas is fed by the winds, which inhibits circularization (see \citealt{Janiuk2008}).

\subsection{Pang et al. 2011}
\cite{Pang2011} (P11) presented 3D MHD simulations in which gas is fed through the outer boundary with uniform magnetic field, spherically symmetric density and pressure, and purely rotational velocity such that the specific angular momentum $l$, varies as $\sin(\theta)$.  This set up is quite similar to PB03B except for the field geometry (uniform vs. radial in PB03B), the field strength, and the addition of a third dimension.  Unlike PB03B, however, P11 found that global magnetic torques and not the MRI were the governing physical mechanisms driving accretion in their simulations. This difference relative to PB03B is probably a consequence of the initial magnetic field in P11 being much stronger, with the initial $\beta$ being $\sim$ $10^2$ in P11 compared to $\beta\sim$ $10^5-10^7$ in PB03B. This causes the magnetic field to become dynamically important before the gas can circularize (if it ever would have) and also suppresses the MRI.  Dissipation of the field also leads to an unstable entropy profile, driving convection. A steady state is reached in which the gas is in near hydrostatic equilibrium, slowly falling inwards with magnetic pressure resisting the upward buoyancy force.  Several aspects of the P11 simulations are similar to what we have found in ours. Both show a lack of circularization, both have the MRI suppressed by strong magnetic fields, and both find a density profile of $\sim$ $r^{-1}$ with a corresponding $\dot M \propto \sqrt{r_{in}}$ relationship.  At the same time, the accretion flow structures are very different in the two sets of calculations.  Unlike P11, the gas in our simulations is not hydrostatic, because the ram pressure, $\rho v^2$ is comparable to or larger than the magnetic and thermal pressures throughout the domain.  We also find significant and coherent outflow, something absent in P11.  

The root cause of these differences are related to the more complicated, asymmetric way that the winds of the WR stars supply gas (and magnetic field) to the black hole.  While both sets of simulations can contain relatively large and coherent magnetic fields at large radii, the steady state of P11 is one in which the gas is being sourced in an approximately spherically symmetric way with rotation playing only a minor role. This is because after the initial transient in which the sourced gas first free falls and then transitions to a PB03A-like configuration, the build up of gas at small radii provides radial pressure support for the gas at large radii, significantly increasing the time it takes to accrete.  At this point the magnetic torques have enough time to remove a large amount of the angular momentum at large radii, ultimately resulting in a quasi-spherical steady-state.  In contrast, because the feeding in our simulations occurs in more of a stream-like manner (Figures \ref{fig:time_series_hydro} and \ref{fig:time_series_beta_1e2}), we have no build-up of gas to provide radial pressure support.  Instead, radial velocities remain large and thus the effect of even significant ($\beta\sim$ few) magnetic torques are limited by the short inflow/outflow times.  This means that rotation of gas is important throughout our simulations, with the distribution of angular momentum being the primary determinant of the accretion rate. 

\section{Implications for Horizon-Scale Modeling}
\label{sec:imp}
The main properties of nearly all GRMHD simulations used to model the Galactic Centre are governed by the evolution of the MRI.  The supply of gas is determined by an initial rotating torus while low angular momentum material is absent.  Our results suggest that for the Galactic Centre it may be critical to consider a more detailed model for how the gas is fed into the domain, particularly with respect to the distribution of angular momentum coming in from larger radii. 

One large remaining uncertainty is how strong the outflows are from near the horizon and whether they significantly modify the dynamics at $\sim$ 1000 $r_g$ found here. Figure \ref{fig:jet_time_series} shows that, at times, we do see strong outflows that can modify the gas out to $\sim 0.3$ pc scales.  Since we find that $\dot M$ $\propto \sqrt{r_{in}}$, the energy released in outflows should scale with the inner boundary as $\propto \dot M v_{kep}^2 \propto r_{in}^{-1/2}$, meaning that the strength of this outflow would be a factor of $\gtrsim \sqrt{150} \approx 10$ times higher if our simulation reached the event horizon.  Additionally, if the black hole is rapidly rotating the magnetic field can extract a significant amount of energy from the black hole and further increase the energy in the outflow \citep{BZ1977}.

The time variability of the polarization vector observed in the GRAVITY flares \citep{GRAVITYFlare} at $\sim $ 10 $r_g$ has been interpreted as the results of an orbitting ``hot spot'' embedded in a face-on rotating flow threaded by a magnetic field primarily in the vertical direction.  Qualitatively, the geometry of the magnetic field at small radii in our $\beta_w=10^2$ simulation agrees with this picture (Figures \ref{fig:fieldlines_comp}), with the poloidal field being larger than $B_\varphi$. On the other hand, the angular momentum direction of the inner accretion flow in Figure \ref{fig:L_time_interval} is rarely as face-on as that of the best-fit orbit of the three flares ($L_z/L\approx 0.94 \pm 0.06$).  

\citet{Psaltis2015} showed that preliminary EHT measurements of the size of the emitting region for Sgr A* 230 GHz are smaller than the expected ``shadow'' of the black hole: the distinct lack of emission caused by the presence of a photon orbit. The authors use this measurement to constrain the angular momentum direction of the disc/black hole (which they assume to be aligned), and find that an inclination angle roughly aligned with the clockwise stellar disc is preferred. This is consistent with measurements of the position angle of the 86 GHz and X-ray emission performed by the VLBA on a scale of $\sim$ 10s of $r_g$ \citep{Bower2014} and by \emph{Chandra} on a much larger scale of $\sim$ 1'' \citep{Wang2013}, respectively.    Our results are in good agreement with these observations, as the angular momentum of our innermost accretion flow is typically aligned with the stellar disc (Figure \ref{fig:L_time_interval}).  Forthcoming higher sensitivity EHT measurements will be important for resolving the discrepancy with the leading interpretation of the GRAVITY data.

\section{Conclusions}
\label{sec:conc}

We have presented the results of 3D simulations of accretion onto the supermassive black hole in the Galactic Centre fueled by magnetized stellar winds. Our simulations span a large radial range, having an outer boundary of 1 pc and an inner boundary of $\sim 6 \times 10^{-5}$ pc ($\sim 300 r_g$), with approximately logarithmic resolution in between.  The mass loss rates, wind speeds, and orbits of the stellar wind source terms that represent the $\sim$ 30 WR stars are largely constrained by observations while the relative strength of the magnetic field in each wind is parameterized by a single parameter $\beta_w$, defined as the ratio between the ram pressure and midplane magnetic pressure of the wind.  In previous work, we have shown that our simulations naturally reproduce many of the observational properties of Sgr A* such as an accretion rate that is much less than the Bondi estimate, a density profile $\tilde \propto$ $r^{-1}$, a total X-ray luminosity consistent with $\emph{Chandra}$ measurements, and the rotation measure of Sgr A*.  In the present paper we have focussed on the dynamics of accretion onto Sgr A* from magnetized stellar winds.

Our most significant and a priori surprising result is that the accretion rate onto the black hole, as well as the radial profiles of mass density, temperature, and velocity are set mostly by hydrodynamic considerations (Figure \ref{fig:mhd_comp}). This is true even when plasma $\beta$ is as low as $\approx$ 2 over a large radial range (Figure \ref{fig:beta_comp}). Without magnetic fields, the accretion rate and density profiles are set by the distribution of angular momentum with accretion rate provided by the stellar winds, a distribution which extends down to $l\approx0$.  This broad range of angular momentum is a consequence of the fact that the WR stellar wind speeds ($\sim$ 1000 km/s) are comparable to their orbital speeds. As a result, the stellar winds provide enough low angular momentum material to result in an extrapolated accretion rate that is in good agreement with previous estimates for Sgr A*.  With magnetic fields, global torques provide only order unity corrections to this picture, with the accretion rate still mostly being determined by the supply of low angular momentum gas.  This is a consequence of the fact that the high angular momentum material in our simulations does not circularize but mostly flows in and out with large enough radial velocity that the inflow/outflow times are short compared to the time scale for magnetic stresses to redistribute angular momentum.  

Simulations with strong magnetic fields at small radii do however differ from hydrodynamic simulations in one important way.  Hydrodynamic simulations are dominated by inflow along the poles, while the midplane is on average outflowing but composed of both inflow and outflow components at different $\theta$ and $\varphi$.  By contrast, MHD simulations are dominated by inflow in the midplane, while the polar regions are on average outflowing but composed of both inflow and outflow components at different $\theta$ and $\varphi$.  This is a consequence of the $\beta \sim$ few magnetic fields redirecting the high angular momentum outflow away from the midplane.

We find that the magnetic field increases rapidly with radius so that $\beta$ tends to eventually saturate at small radii to a value of order unity independent of $\beta_w$ (Figure \ref{fig:beta_comp}).  
This growth of the field is caused by advection/compression as gas falls inwards and not by the MRI. There is neither sufficient time for the MRI to grow before gas is accreted or advected to larger radii, nor is there sufficient space for the instability to grow because flux freezing builds up a field for which the most unstable MRI wavelength is comparable to or larger than the disc scale height (Figure \ref{fig:lambda_mri}).  Thus the conventional MRI-driven torus simulations that dominate the literature do not appear to have reasonable initial conditions for studying accretion in the Galactic Centre, at least on the scales that we can simulate here.  

Elaborating on the result first presented RQS19, we have shown that our model predicts that the magnetic flux ultimately threading the event horizon, $\phi_{in}$, will be on the order of 5, independent of $\beta_w$ for $\beta_w \lesssim 10^7$ (Figure \ref{fig:phi_in_rin}).  This prediction relies on extrapolation to smaller radii, ignores the effects of GR, and assumes that the scaling between $\phi_{in}$ and the inner boundary radius that we found (Figure \ref{fig:phi_in_rin}) holds at smaller radii than our simulations probe. Not accounting for GR effects, this amount of flux threading the horizon is potentially large enough to induce a MAD state near the horizon, with the outward Lorentz force reaching $\lesssim$ 10\% of the inward gravitational force in our simulations (Figure \ref{fig:press_radius}), which is $\gtrsim$ the ram pressure due to $v_r$.  It is worth noting that our lower $\beta_w$ simulations do in fact display many similar properties to GRMHD MAD accretion flows, for instance, the MRI is suppressed by strong vertical fields, the poloidal component of the field dominates over the toroidal component, and the angular velocity of the gas is roughly half the Keplerian value \citep{Narayan2012}, though we find the latter to be true in both MHD and hydrodynamic simulations. We also find that $\phi_{in}$ is relatively independent of time (Figure \ref{fig:phibh_comp}).  There are, however, a number of ways in which the simulations in the present work do not appear to be fully magnetically arrested.   For example, the hydrodynamic and MHD simulations show similar accretion rates and radial profiles (Figures \ref{fig:mhd_comp} and \ref{fig:mdot_beta_comp}), which explicitly demonstrates that the magnetic field is not dynamically critical for establishing the flow properties.   In addition, our simulations find very different angular distributions of density (Figure \ref{fig:rho_theta}) and plasma $\beta$ from traditional MAD simulations (and, in fact, most GRMHD simulations). Because of the significant amount of low angular momentum material provided by the stellar winds, the polar regions are only a factor or 3--5 less dense and have only a slightly lower $\beta$ than the midplane. In contrast, a typical GRMHD MAD simulation would show near-vacuum, highly magnetized polar regions with orders of magnitude less mass than in the midplane, where most of the mass is condensed into a relatively thin disc compressed by the magnetic field.  The evacuation of the funnel is caused by both the strong jets present in MADs \citep{Igumenshchev2003,Sasha2011} as well as the fact that the rotating hydrodynamic configurations used as initial conditions in GRMHD simulations generally do not allow material close to the poles \citep{Abramowicz1978,Kozlowski1978}. We do not at all, however, rule out that by the time the gas reaches the event horizon of the black hole that a MAD state could be reached. This may also depend on whether or not the black hole is rotating, since a rotating hole provides an additional source of energy for outflows that could strongly impact the dynamics in the polar region.

For sufficiently magnetized winds (i.e., $\beta_w=10^2$ here), the magnetically driven, polar outflow can, at times, reach scales as large as $\sim$ 0.3 pc (Figure \ref{fig:jet_time_series}).  Since we expect the energy associated with this outflow to increase with decreasing inner boundary radius, it could potentially be a factor of $> 10$ times stronger in a simulation that reached the event horizon. This is even without considering the rotation of the black hole itself, which can also be an efficient mechanism for driving magnetized jets \citep{Hawley2006}. Though there is no clear signature of a jet in the Galactic Centre, strong outflows from Sgr A* have been invoked as one possible explanation for the recent ALMA observations that show highly blue-shifted emission from unbound gas in a narrow cone \citep{Royster2019}. 

The magnetic field structure at small radii depends on the parameter $\beta_w$ (Figure \ref{fig:fieldlines_comp}).  For smaller $\beta_w$ ($10^2$ and to a lesser extent, $10^4$) the field is strong enough to resist being wound up in the $\varphi$ direction and remains mostly poloidal at small radii. For larger $\beta_w$ ($\gtrsim 10^6$), the field is easily dragged along with the motion of the gas so that it becomes predominantly toroidal by the time $\beta$ reaches order unity.  The leading interpretation of the GRAVITY observations of astrometric motion of the IR emission during Sgr A* flares \citep{GRAVITYFlare} requires that the horizon scale magnetic field be mostly perpendicular to the angular momentum of the gas.  We find qualitative agreement with this result in our simulations that have more magnetized winds. A more quantitative comparison to the observations using full radiative transfer in GRMHD simulations using such a field as initial conditions will require additional work.

\citet{Cuadra2008} found that the winds of only 3 WR stars (E20/IRS 16C, E23/IRS 16SW, and E39/IRS 16NE)  dominated the $t=0$ accretion budget in their simulations that used the `{\tt 1DISC}' orbital configuration.  This is both because of the proximity and relatively slow wind speeds ($\sim$ 600 km/s) of these winds.  Since we adopted the `{\tt 1DISC}' configuration from \citet{Cuadra2008} with only slight changes, it is not surprising that the same three stellar winds seem to be the most important for determining the properties of the innermost accretion flow in our simulations\footnote{Unlike the particle based calculation of \citet{Cuadra2008}, we do not have a rigorous way to track the gas from each individual wind in our current implementation.  We can only infer which stellar winds dominate the accretion budget from, e.g., the poloidal and toroidal animations.} (e.g., Figures \ref{fig:time_series_hydro} and \ref{fig:time_series_beta_1e2}). Future observations that place stronger constraints on the mass-loss rates, wind speeds, and especially the magnetic field strengths of these stars would thus go a long way towards reducing the uncertainty in our calculation.

Several observations suggest that gas surrounding Sgr A* is aligned with the clockwise stellar disc both near the horizon and just inside the Bondi radius (\citealt{Wang2013,Bower2014,Psaltis2015}; though see also \citealt{GRAVITYFlare}).  Our simulations are consistent with this result for $\beta_w\ge 10^2$ (Figure \ref{fig:L_time_interval}) but not for $\beta_w = 10$ due to wind collimation altering the distribution of angular momentum in the winds (not shown).  If a large fraction of the accreting gas (and associated magnetic field) were sourced from material outside the region where the majority of the WR stars reside, then it would also be unlikely for its angular momentum to coincide with the stellar disc.  In all of our simulations, the direction of the angular momentum of the inner accretion flow is not strictly constant in time over the $\sim$ 1000 yr duration of our simulation (Figure \ref{fig:L_time_interval}).  Therefore, the angular momentum of the gas sourcing the horizon scale accretion flow \emph{must} be tilted with respect to the spin of the black hole at least moderately often since the time scale for the spin of the black hole to change is much longer than 1000 yr. Simulations of tilted accretion discs (like those of \citealt{Fragile2005,Liska2017,Hawley2019}) are thus likely necessary for horizon scale modeling of Sgr A*.

Our results could have a significant impact on current state of the art models of horizon scale accretion onto Sgr A*.  GRMHD simulations to date almost universally rely on the MRI as the mechanism to drive accretion.  It is not clear how much the results of these simulations and their observational consequences might change using the dynamically different flow structure found here.  For instance, if the disc is less turbulent without the MRI, how does this effect the time-variability properties of the emission? Would nearly empty, magnetically dominated jets still be robustly present in GRMHD and does this depend on black hole spin and horizon-scale flux in the same way as in current simulations (e.g. \citealt{Sasha2011})?  Such questions and more will be important to answer in order to further our understanding of the emission from Sgr A* and other low luminosity AGN.

\section*{Acknowledgments}
We thank the referee, Ramesh Narayan for a careful and thorough reading of this manuscript.  We thank F. Yusef-Zadeh for useful discussions, as well as all the members of the horizon collaboration, \href{http://horizon.astro.illinois.edu}{http://horizon.astro.illinois.edu}. SR was supported by the Gordon and Betty Moore Foundation through Grant GBMF7392 for part of the duration of this work.  EQ thanks the Princeton Astrophysical Sciences department and the theoretical astrophysics group and Moore Distinguished Scholar program at Caltech for their hospitality and support. This work was supported in part by NSF grants AST 13-33612, AST 1715054, AST-1715277, \emph{Chandra} theory grant TM7-18006X from the Smithsonian Institution, a Simons Investigator award from the Simons Foundation, and by the NSF
through an XSEDE computational time allocation TG-AST090038
on SDSC Comet.  This work was made possible by computing time granted by UCB on the Savio cluster.

\bibliographystyle{mn2efix}
\bibliography{mhd_wind}

\appendix

\section{Resolution Study}
\label{app:resolution}
\begin{figure}
\includegraphics[width=0.45\textwidth]{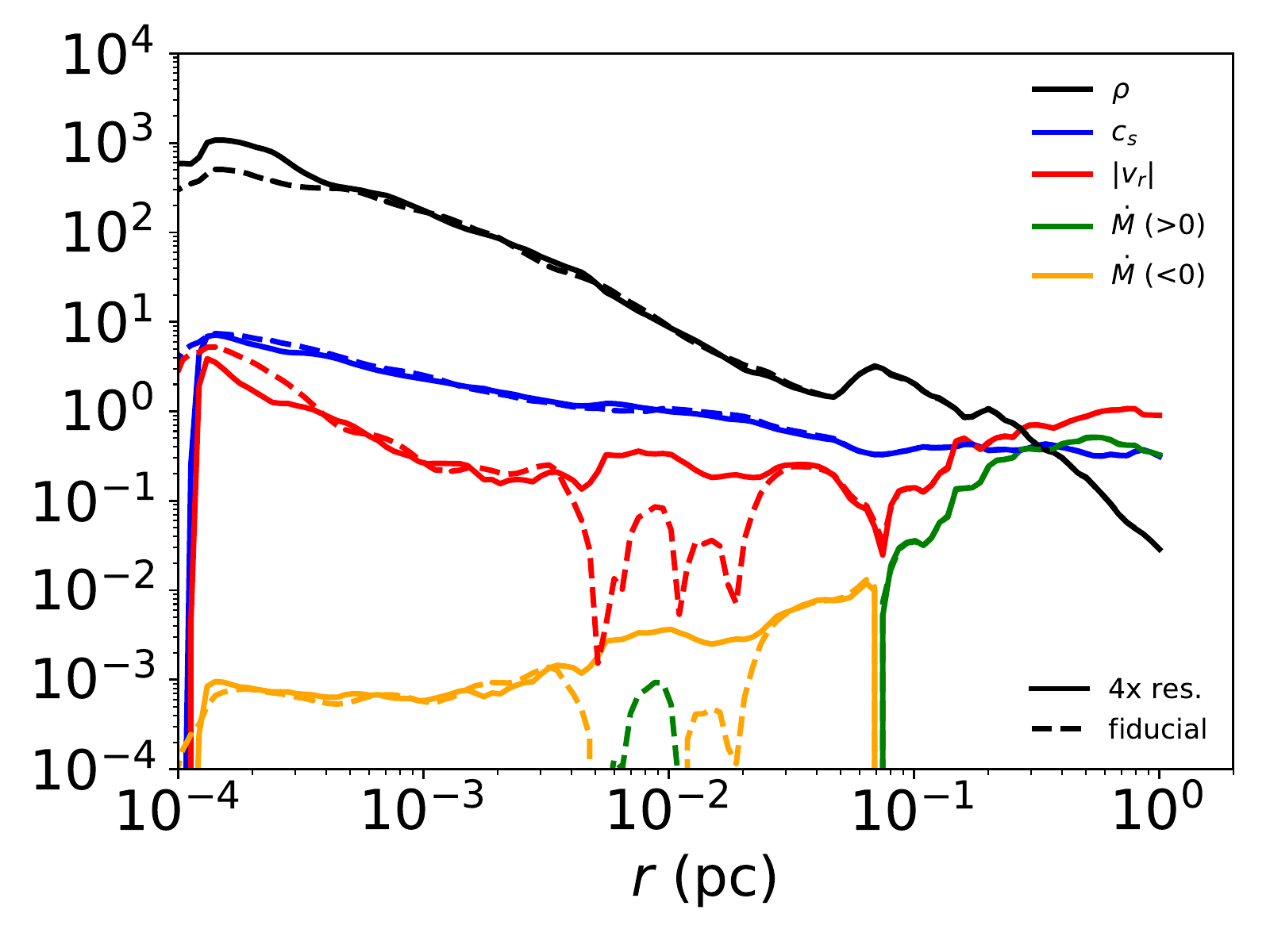}
\caption{Demonstration of convergence for $\beta_w=10$ simulations.  Dashed lines are from a simulation at the fiducial resolution while solid lines are from a simulation with a factor of 4 higher resolution for $r\lesssim$ 0.06 pc.  The angle averaged mass density, $\rho$ (black), sound speed, $c_s$ (blue), radial velocity, $v_r$ (red), and net accretion rate, $\dot M$ (green for positive, orange for negative), are essentially identical after being run for 25 yr (one orbital time at $\approx 0.007$ pc) and thus converged at the fiducial resolution used throughout this work. 
}
\label{fig:res_comp}
\end{figure}

We have argued in the main text in \S \ref{sec:MRI} that the MRI is not the governing mechanism for accretion in our simulations even though the most unstable wavelength is well resolved (Figure \ref{fig:lambda_mri}).  However, to be assured that resolution is not affecting our results, we performed two additional simulation with $\beta_w=10$ and an inner boundary radius of $r_{in} \approx 1.2 \times 10^{-4}$ pc.  The first simulation was run for 1.025 kyr until $t=-0.075$ kyr with our usual base resolution of 128$^3$ cells with 8 levels of mesh refinement that increase the resolution by a factor of 2 each time the radius decreases by a factor of 2.   The second simulation increased the resolution by a factor of four within $\sim$ 0.06 pc and ran for 25 yr after being restarted from the lower resolution simulation at -0.1 kyr. 25 yr is approximately an orbital time at $0.007$ pc and thus spans many orbital times for the small radii of interest.  Both calculations were done without radiative cooling in order to reduce the computational cost.   Figure \ref{fig:res_comp} shows that the resulting radial profiles of angle-averaged fluid quantities in the two simulations are nearly identical.  Thus we are confident that the general properties of our simulations are well converged and not limited by resolution.

\section{Simulations With Longer Run Times}
\label{app:t0_comp}
In this Appendix we show that our results are not sensitive to the start time ($t_0$) of our simulations, that is, the the length of time we run our simulations before the present day at $t=0$.  In principle, $t_0$ should be chosen to represent the typical duration of the WR phase ($\sim 100$ kyr), however, such a simulation would be extremely expensive and perhaps unnecessarily so if the resulting dynamics of the accretion flow are mostly sensitive to the $t=0$ location of the WR stars and not the accretion history.   Previous work (e.g., \citealt{Cuadra2008}, R18, R19) has argued that a run time of $\sim$ 1.1 kyr is sufficient to study accretion at $t=0$.   As this manuscript was in press, however, \citet{Calderon2019} (C19) released new results of conservative, grid-based, hydrodynamic simulations of accretion via stellar winds onto Sgr A* that suggest otherwise.  Using the {\tt RAMSES} code \citep{RAMSES} with the same wind speeds, orbits, and mass-loss rates of the WR stars used in the present work, they studied the effect of starting their simulations further back in time.  The results of their control run with $t_0=-1.1$ kyr (as used in the main text of this work) were generally consistent with ours.  In their $t_0=-3.5$ kyr simulation, however, they found that a cold, dense disc of gas formed after $\sim$ 2 kyr that significantly increased the accretion rate through the inner boundary and altered the radial profiles of density and temperature at $t=0$.  Presumably the longer the simulation is run the more massive the disc becomes.  Based on this finding, they conclude that simulations using $t_0=-1.1$ kyr (including \citealt{Cuadra2008}, R18, R19, and now this work) are modeling a `quasi-steady state' that may not be representative of the current accretion flow around Sgr A*.  

\begin{figure}
\includegraphics[width=0.45\textwidth]{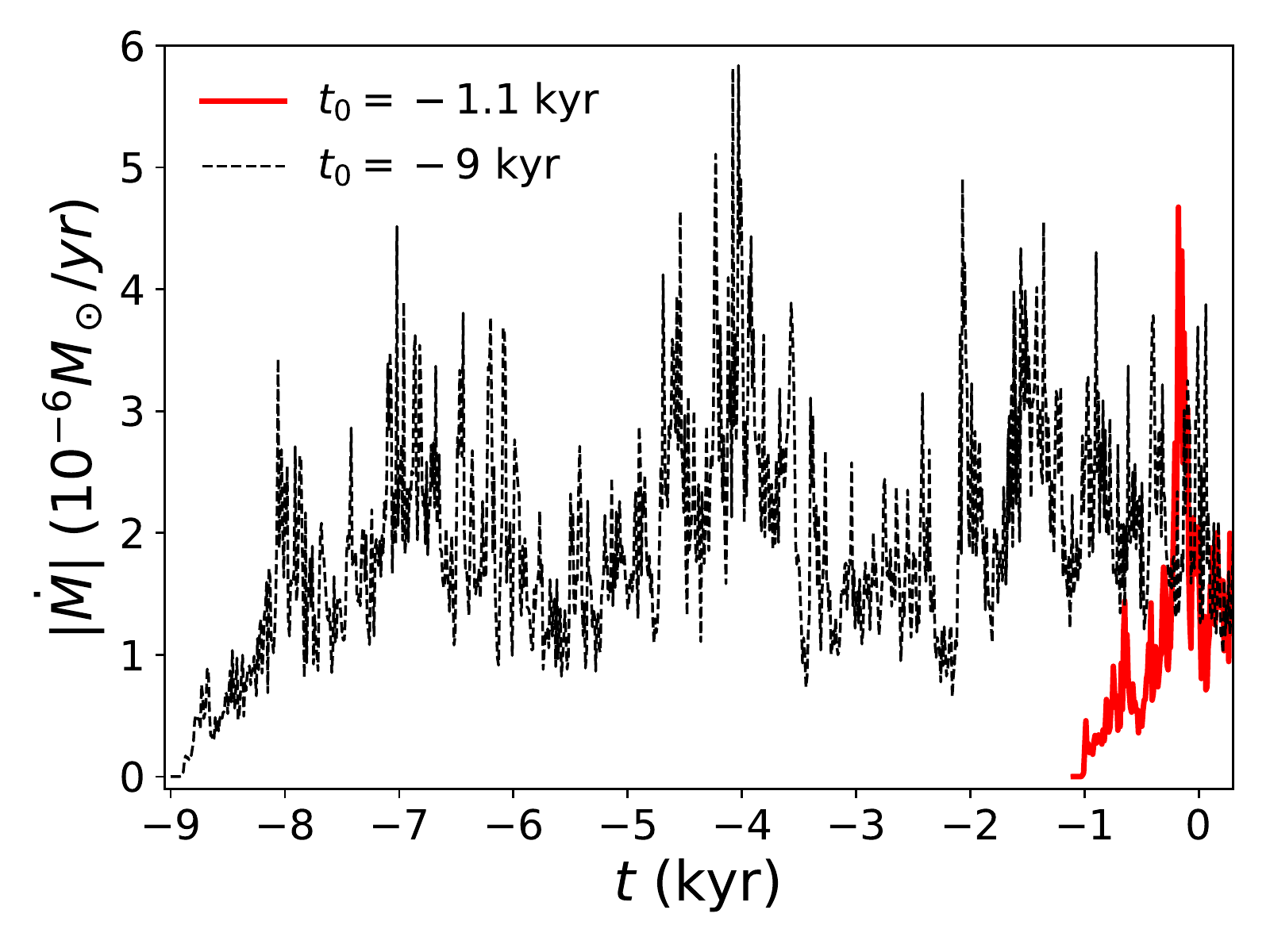}

\caption{Accretion rate vs. time at $r\approx 5 \times 10^{-4}$ pc for hydrodynamic simulations with $t_0 = -1.1$ kyr (solid) and $t_0=-9$ kyr (dashed).    The longer run time (i.e., earlier start time)  has a negligible effect on the resulting accretion rate around $t=0$. This demonstrates that $t_0$ = -1.1 kyr is sufficiently far back in time before the present day to start our simulations.}
\label{fig:t0_comp}
\end{figure}

In light of this result, we revisited the question of how sensitive our results are to the choice of $t_0$ by running an additional $t_0=-9$ kyr simulation in hydrodynamics.  By reaching a final time of $t_f = 0.2$ kyr, this simulation is thus run $\sim$ 9 times longer than the simulations described in the main text and $\sim$ 2.6 times longer than the $t_0=-3.5$ kyr simulations of C19.  To make the computational cost more manageable, we use only 7 levels of nested refinement, corresponding to $r_{in} \approx 2 \times 10^{-4}$ pc.  Figure \ref{fig:t0_comp} shows the accretion rates through $r \approx 5 \times 10^{-4}$ pc for the $t_0=-9$ kyr simulation compared to the hydrodynamic simulation with $t_0=-1.1$ kyr.  Near $t=0$ the $t_0=-1.1$ kyr and $t_0=-9$ kyr simulations display remarkably similar accretion rates, suggesting that the $t=0$ dynamics are largely determined by the present day locations and velocities of the WR stars.  We find this to be true also of the radial profiles of gas quantities.  In Figure \ref{fig:t0_comp}, the $t_0=-9$ kyr simulations show no evidence for the `disc phase'  seen in C19, nor do we find any build up of mass with time.   Thus, we find that the dynamical picture outlined in the main text (such as the lack of circularization and lack of disc formation) is still valid for simulations run a factor of 9 times longer than our fiducial model.  We reach similar conclusions from MHD simulations run with $t_0$ further back in time than -1.1 kyr. Though we cannot run a simulation going as far back as the $\sim$ -100 kyr timescale appropriate for the lifetime of a typical WR star, we would not expect to find any significant differences from the $t_0=-9$  kyr and $t_0=-1.1$ kyr simulations.

Since the physics and numerical methods of our models and those of C19 are similar (at least for our hydrodynamic calculations), it is not clear what causes such a striking discrepancy in results.  The biggest difference in our implementation seems to be the specific treatment of the stellar winds:  C19 use the technique described in \citet{Lemaster2007} that overwrites certain cells with the analytic solution of a spherical wind while we treat the winds as source terms spread over a certain number of cells (see \S \ref{sec:Model}) that ultimately drive a spherical wind.  
Overall, our resolution is a factor of $\sim$ 2  higher than C19, though they use adaptive mesh refinement to place finer grids in the stellar wind regions giving them better local resolution where these winds are being generated.  That said, it is not obvious which (if any) of these seemingly small numerical details might be linked to the ensuing formation or absence of a cold disc; this will be an important issue to resolve going forward.

\end{document}